\documentclass[pre,aps,twocolumn,superscriptaddress,floatfix]{revtex4-1}
\usepackage[english]{babel}
\usepackage[dvipsnames]{xcolor}
\usepackage{amsmath} 
\usepackage{color}
\usepackage{graphicx}
\usepackage{dcolumn}
\usepackage{bm}
\usepackage[colorlinks=true, linkcolor=blue, citecolor=blue, urlcolor=blue]{hyperref}
\usepackage{float} 
\usepackage{xcolor}

\newif\ifmarked

\ifmarked
  \colorlet{myblue}{blue}
\else
  \colorlet{myblue}{black}
\fi

\begin{document}

 \title{
Geometric flow of 
planar domain-wall loops 
 }

\author{P. Domenichini}
\affiliation{ INENCO, CONICET. Salta Capital, Salta, Argentina}
\affiliation{ Universidad Nacional de Salta, Departamento de F\'{\i}sica. Salta Capital, Salta, Argentina}

\author{G. Salazar}
\affiliation{ INENCO, CONICET. Salta Capital, Salta, Argentina}
\affiliation{ Universidad Nacional de Salta, Departamento de F\'{\i}sica. Salta Capital, Salta, Argentina}

\author{A. B. Kolton}
\affiliation{Centro At\'omico Bariloche, CNEA, CONICET, Bariloche, Argentina}
\affiliation{ Instituto Balseiro, Universidad Nacional de Cuyo, Bariloche, Argentina}

\date{\today}

\begin{abstract}
We investigate the geometric  evolution of elastic domain-wall loops in two dimensions. Assuming an instantaneous, isotropic, and homogeneous arc-velocity response of the domain wall to external pressure and local signed curvature, we derive closed dynamical equations linking the enclosed area and loop perimeter for both linear and nonlinear arc-velocity response functions.
This reduced description enables predictions for the dynamics of both spontaneous and externally driven domains—subjected to constant or alternating fields—within the time-dependent Ginzburg–Landau scalar $\phi^4$ model. In the linear response regime, where an avoidance principle holds in the absence of external driving, we obtain exact results. In particular, we demonstrate that the relaxation rate of the total spontaneous magnetization becomes quantized for arbitrary initial conditions involving multiple, possibly nested, loops, with discrete jumps corresponding to individual loop collapse events.
Under external driving, the avoidance principle breaks down due to sparse interactions between interfaces—either within a single loop or between multiple loops—leading to coalescence or splitting events that change the number of loops. A quantized geometrical observable involving the total area and perimeter is identified in this case as well, exhibiting discrete jumps both at interface interaction events and at individual loop collapses.
We further use approximate area-perimeter relations to estimate the spontaneous collapse lifetimes of compact magnetic domains, as well as their dynamics under alternating-field-assisted collapse in disordered ultrathin magnetic films. Our predictions are compared with experimental observations in such systems.
\end{abstract}

\pacs{Valid PACS appear here}

\maketitle

\section{Introduction}
\label{sec:intro}

Planar elastic loops, whose energy is proportional to their perimeter, are ubiquitous, as they can arise as domain walls enclosing compact regions of certain order parameters in two-dimensional systems.
A paradigmatic physical example is the domain wall enclosing a compact magnetic domain in an isotropic two-dimensional ferromagnetic material with uniaxial anisotropy, nucleated within a uniformly magnetized background \cite{Quinteros2020, lemerle1998domain, chauve2000creep}.
In such systems, a uniform external magnetic field favors the growth of the domain area, while the domain wall energy, proportional to its perimeter, tends to drive the collapse of the domain.
It is therefore natural to ask whether the domain’s evolution can be captured by reduced dynamical equations involving only two global geometric quantities: area and perimeter as functions of time \cite{RevModPhys.90.045006, kardar1998nonequilibrium}.
A particularly suitable experimental realization is provided by magnetic domains in ultrathin ferromagnetic films , which can be nucleated with a magnetic pulse, driven by smaller external fields, and imaged in real time using magneto-optical techniques 
\cite{ferre2013universal}.

A domain-wall loop enclosing a compact domain is an extended system. As such, a complete description of its dynamics using only a small number of variables is generally not feasible, since infinitely many distinct shapes—with different perimeters—can enclose the same area. 
However, in certain physical contexts, a reduced dynamical description becomes possible, drawing inspiration from the well-known curve-shortening flow problem in mathematics. For instance, a simple closed planar curve, whose local normal velocity is proportional to its local curvature, contracts its enclosed area at a constant rate, independent of the initial shape.
If the curve is initially simple (i.e., it does not intersect itself), an avoidance principle ensures it remains simple throughout its evolution. Over time, the curve becomes convex and ultimately circular before collapsing to a point as rigorously described by the Gage–Hamilton–Grayson theorem for simple curves ~\cite{GageHamilton1986,grayson1987heat,white2002evolution}.

In this paper, we adapt some of these well-established mathematical results to the physical context of domain-wall dynamics in two dimensions, employing suitable approximations that we critically evaluate. We investigate the effects of constant and oscillating external fields, quenched disorder, thermal fluctuations, nonlinearities in the local arc-velocity response of the domain wall, and the effect of short-range interactions between interfaces with a finite-width. We show that meaningful approximations for the domain evolution can be derived. To assess the accuracy of these approximations, we compare their predictions with numerical simulations of a universal isotropic model for finite-width domain-walls: the $\phi^4$ model. Finally, we illustrate the practical relevance of our results by making predictions and interpreting quantitatively experimental results previously obtained  in ultrathin magnetic systems.
\color{myblue}
\section{Geometrical Framework and Notation}
\color{black}
Let us consider the dynamics of a closed domain wall in a plane, described at time $t$ by the smooth curve $\Gamma _t$, and parameterized by $s$.
We assume that each differential arc-element of length $\mathrm{d}s$ at position ${\bf r}_s$
has an overdamped dynamics and responds instantaneously to all the forces acting on it, such that its instantaneous velocity $v_s$ in the local normal direction ${\hat n}_s$ is given by
\begin{figure}[htbp]
    \centering
    \includegraphics[width=\linewidth]{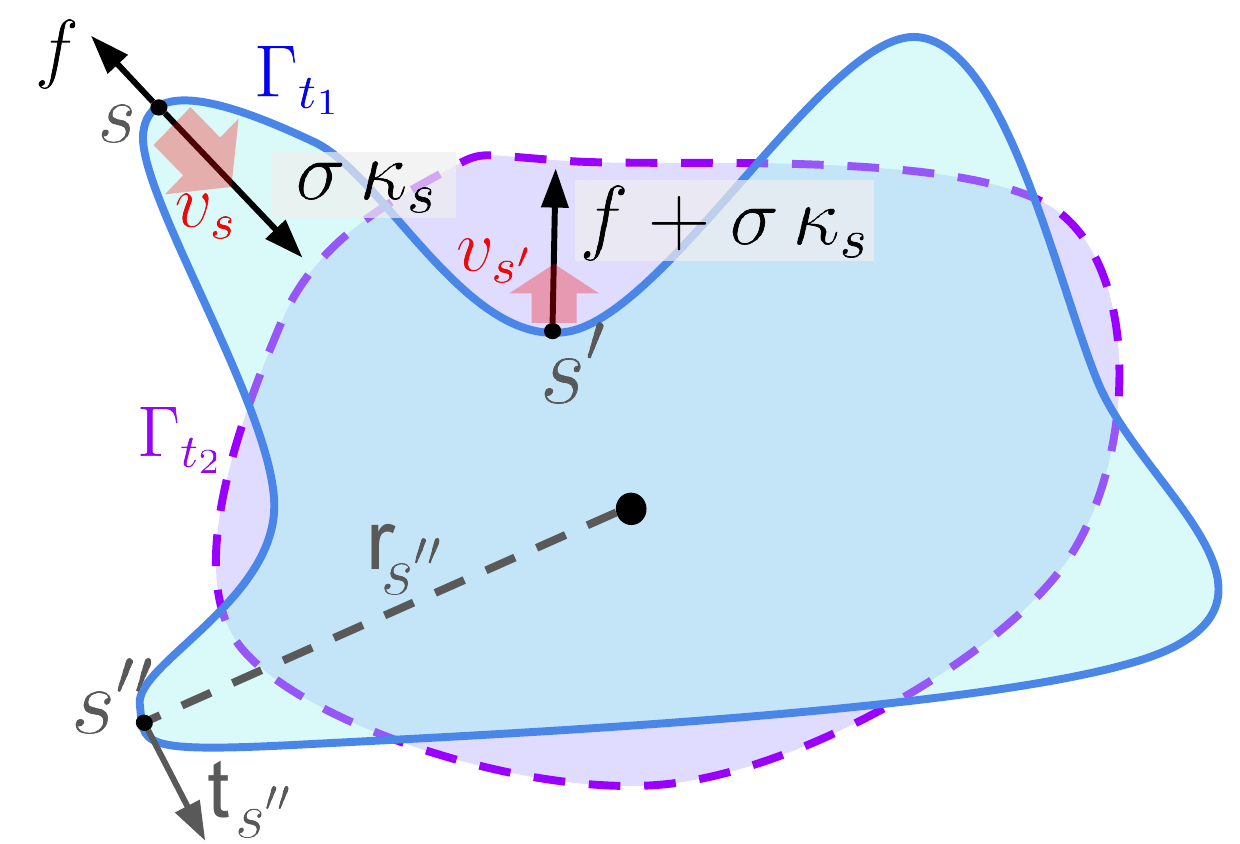}
    \caption{Schematic of an elastic DW loop evolving in time under signed curvature forces, $\sigma \kappa_s$, and an external local pressure $f$. The local normal velocity $v_s$ follows a general response function $V$ of these two forces, as given in Eq.~\eqref{eq:arcdynamics}. The loop is described not only by its enclosed area $A(t)$ and perimeter $P(t)$, but also by the orientation of its tangent vector ${\hat t}_s$ along the contour. By convention, we take ${\hat t}_s$ anticlockwise when $f>0$ produces domain growth.}
    \label{fig:scheme}
\end{figure}

\begin{align}
    v_s  = V(\sigma\kappa_s + f) 
    \label{eq:arcdynamics}
\end{align}
where $\sigma$ is a surface tension coefficient, $\kappa_s$ the instantaneous local signed curvature  computed from $\Gamma_t$ at $s$, $f$ an applied possibly time-dependent but uniform field pressure, and $V$ is a smooth and homogeneous arc-velocity response function such that $V(0)=0$ (see Fig.\ref{fig:scheme}).
Using a tangent vector $\hat t_s$ we assign an orientation to the loop with the convention that when $f>0$ and anticlockwise orientation produces domain growth (therefore  $\kappa_s < 0$ for convex domains with anticlockwise orientation).

To evaluate Eq.~(\ref{eq:arcdynamics}), we require an explicit parametrization of the instantaneous curve $\Gamma_t$ describing the loop at each time. However, if $\Gamma_t$ is a simple (i.e. non-self-intersecting) planar closed-curve-- also known as Jordan curve-- the time evolutions of its enclosed area $A \equiv \frac{1}{2} |\int_{\Gamma_t} (x(s)\, dy(s) - y(s)\, dx(s))|$ and its perimeter $P\equiv \int_{\Gamma_t} ds$ are \color{myblue}
exactly described by \cite{white2002evolution}
\color{black}
\begin{eqnarray}
\frac{dA}{dt} &=& \int_{\Gamma_t} v_s ds \label{eq:dAdt} \\
\frac{dP}{dt}&=&-\int_{\Gamma_t} v_s \kappa_s ds
\label{eq:dPdt}
\end{eqnarray}
These equations provide exact expressions for the evolution of $A$ and $P$, but they do not constitute a closed dynamical description, as evaluating $v_s$ requires knowledge of the full curve evolution $\Gamma_t$ through Eq.~\eqref{eq:arcdynamics}.

Eqs~\eqref{eq:dAdt} and \eqref{eq:dPdt} hold generally for simple, closed, smooth curves with local normal arc-velocities $v_s$, and are independent of the specific form of the arc-velocity response function $V(x)$ in Eq.~\eqref{eq:arcdynamics}, where $x = f + \sigma \kappa_s$ represents the local (2D) pressure.
We will discuss two cases separately. The first, and simpler, case corresponds to an isotropic and linear arc-velocity response, $V(x) \propto x$, corresponding to the Allen–Cahn-type domain wall dynamics~\cite{Bray1994}. This scenario is physically motivated by the expected overdamped motion of magnetic domain walls in a homogeneous and isotropic medium, provided we operate within a regime where the internal degrees of freedom of the domain wall are negligible. The second, more complex case involves a general nonlinear response function $V(x)$. This can be physically motivated by the nonlinear velocity-field characteristics observed in magnetic domain walls within a homogeneous medium, particularly when internal degrees of freedom become relevant—leading, for example, to phenomena such as Walker breakdown \cite{Schryer1974} within the considered range. Interestingly, the second case can also be viewed as a homogeneous approximation of the coarse-grained arc dynamics of a magnetic domain wall in a disordered medium, by assuming that 
$V(f + \sigma \kappa_s) \approx V_T\bigl(H + (\sigma / 2 M_{\rm s}) \kappa_s \bigr)$, 
where $V_T(H)$ is the experimentally accessible velocity-field characteristic of domain walls under a uniform external magnetic field $H$ at temperature $T$, and $M_{\rm s}$ is the saturation magnetization. In this scenario, the function $V_T(H)$ is a nontrivial and nonlinear function in the depinning and creep regimes, exhibiting universal features.

\section{Elastic loops in the $\phi^4$ Model}
\label{sec:phi4model}
For the numerical computations, we consider the paradigmatic time-dependent Ginzburg--Landau scalar $\phi^4$ model in two dimensions. This model describes the dynamics of a non-conserved scalar order parameter $\phi(x,y)$, which we will refer to simply as the ``magnetization'' (although it is not necessary to restrict the interpretation to domain walls in magnetic systems).  
The general form of the equation of motion we consider is \cite{kolton2023, Caballero2018, guruciaga2019ginzburglandau}
\begin{equation}
\gamma \, \partial_t \phi = c \nabla^2 \phi + 
\epsilon_0\left[(1 + r(x,y)) \phi - \phi^3\right] + h + \xi(x,y,t),
\label{eq:phi4}
\end{equation}
where $\gamma$ is a friction constant, $c$ is an elastic constant, $\epsilon_0$ sets the average height of the double-well barrier, $h$ is a uniform and possibly time-dependent external field, and $\xi\equiv \xi(x,y,t)$ is a thermal noise term with the standard statistical properties:
$
\langle \xi(x,y,t) \rangle = 0, \langle \xi(x,y,t)\, \xi(x',y',t') \rangle = \frac{2T}{\gamma} \delta(x - x')\, \delta(y - y')\, \delta(t - t').
$
These noise correlations ensure thermal equilibrium at sufficiently long times. We also include a quenched disorder function $r\equiv r(x,y)$ to model heterogeneous media.

Within the $\phi^4$ model, the total magnetization of a sample is defined as $M \equiv \int_{x,y} \phi$. The dynamics of compact domains can be studied by initializing the field with a configuration such that $\phi(x,y,t=0) > 0$ for $(x,y)$ belonging to a compact domain of area $A(t=0)$, and $\phi(x,y,t=0) < 0$ elsewhere.
In the absence of disorder and thermal fluctuations, $\phi(x,y,t)$ rapidly relaxes to a configuration with a domain wall loop separating regions of negative magnetization $\phi \approx -\phi_{\rm s}$ and positive magnetization $\phi \approx \phi_{\rm s}$, where $\phi_{\rm s} > 0$ is the saturation magnetization. 
In contrast to micromagnetics, where the saturation magnetization $M_{\rm s}$ is a fixed constraint ($|\mathbf{M}|=M_{\rm s}$), the $\phi^4$ scalar model allows for a weak $h$-dependence of $\phi_{\rm s}$ (see Appendix \ref{Phi4-Details}).

The initial condition with a compact domain in a saturated background yields a magnetization governed by the domain area $A(t)$, such that
\begin{equation}
M(t) \approx [A(t) - A_{\rm sample}]\, \phi_{\rm s},
\end{equation}
where $A_{\rm sample}$ is the constant total area of the sample. The closed domain wall, with perimeter $P(t)$, can be instantaneously located at the zero level set $\Gamma_t$ of $\phi$; that is, the set of points $(x_t, y_t)$ such that $\phi(x_t, y_t, t) = 0$. The typical width of the wall is $\delta \approx \sqrt{c/\epsilon_0}$.
In most cases, the intrinsic profile of the domain wall along its normal direction is approximately rigid, in the sense that spatial variations of the wall profile can be neglected compared to changes in the domain wall shape, which are associated with transverse modes.


We can interpret Eq.~(\ref{eq:phi4}) as a type-A relaxational dynamics \cite{hohenberg1977},
$
\gamma \partial_t \phi = -\frac{\delta \mathcal{F}}{\delta \phi} + \xi,
$ where $\mathcal{F} \equiv \mathcal{F}[\phi]$ 
\color{myblue}
is a free energy functional \cite{1977RvMP...49..435H}. \color{black}
In the absence of quenched disorder (i.e., for constant $r$), the total free energy excess $\Delta F$ per unit length associated with a single compact domain in a saturated background opposing the external field $h$ is approximately,  
\begin{align}
\Delta F \approx \sigma P - 2 \phi_{\rm s} h A 
\label{eq:fdw}
\end{align}
where $\sigma \approx \sqrt{c \epsilon_0}$ is the domain wall surface tension, $P$ is the perimeter of the domain wall loop, $A$ is the enclosed domain area (which grows when $h > 0$)
.
In writing Eq. \eqref{eq:fdw} we have neglected small $\mathcal{O}(2 \phi_{\rm s} P \delta)$ corrections.

\color{myblue}
From Eqs.~\eqref{eq:arcdynamics}, and \eqref{eq:fdw}, 
\color{black}
one can check  
that $f=2\phi_{\rm s} h$ and, in the absence of disorder and thermal fluctuations, that the DW friction coefficient is the constant $\eta \propto \gamma/\delta$~\cite{Caballero2020b,kolton2023} at any point $\mathbf{r}_s$ of the instantaneous curve $\Gamma_t$ at a given instant $t$. The DW friction constant is related to the so-called DW mobility, $m \equiv dV/dh \equiv 2\phi_{\rm s}/\eta$.
In connecting the interface dynamics with that of $\phi$, we are assuming that the characteristic length scale of DW distortions is small compared with $\delta$, i.e., $\kappa_s \ll 1/\delta$.

To solve Eq.~(\ref{eq:phi4}), we discretize the system on a regular $L \times L$ grid with spatial indices $(i,j)$ and discrete time steps indexed by $n$. The discrete evolution equation reads
\begin{eqnarray}
\label{eq:discrete}
\frac{\phi^{n+1}_{i,j} - \phi^{n}_{i,j}}{\Delta t} 
&=& c \bigl(\phi^n_{i+1,j} + \phi^n_{i-1,j} + \phi^n_{i,j+1} + \phi^n_{i,j-1} - 4 \phi^n_{i,j} \bigr) \nonumber \\
&+& \epsilon_0 \bigl[(1 + r_{i,j}) \phi^n_{i,j} - (\phi^n_{i,j})^3 \bigr] + h^n + \xi^n_{i,j}.
\end{eqnarray}
This system is then solved iteratively using explicit Euler time stepping with a suitably small time step $\Delta t$. Our numerical protocol consists of initializing the simulation with a given number of compact domains and then tracking their dynamics and geometric properties over time. The simulations are accelerated by parallelization in graphics processing units (GPUs) using standard libraries.

\section{Summary of main results and Outline}

Our aim is to explore the utility of Eq. \eqref{eq:arcdynamics} for modelling domain-wall loops in increasingly complex situations.
To do so we combine analytical arguments, numerical simulations of a scalar-field model, and the analysis of experimental data, following a progression from simple to increasingly complex situations.
Our main results can be summarized as follows:
\begin{itemize}
    \item \textbf{Curve-shortening flow of physical domain-wall loops}: 
    We show that theorems associated with curve-shortening flow for elastic curves are accurately satisfied by finite-width domain-wall loops in the paradigmatic time-dependent $\phi^4$ scalar model in homogeneous media.

    \item \textbf{Quantization of the total magnetization rate in systems of loops}: 
    For systems comprising multiple domain-wall loops (or compact domains) with a linear and homogeneous arc-velocity response, such as the homogeneous $\phi^4$ model, we show that curve-shortening flow, together with interaction events under an external field (loop collapse or loop merging), implies the quantization of the rate of change of the total magnetization.

    \item \textbf{Extension to nonlinear velocity responses}: 
    We demonstrate that dynamical geometric relations between the loop perimeter and the enclosed area can be extended in a practical form, 
    from Eq. \eqref{eq:arcdynamics} with $V$ a generic non-linear function, to systems with a generic \textit{nonlinear} homogeneous arc-velocity response which may arise in more complex media.

    \item \textbf{Testing the nonlinear extension}: 
    We test the nonlinear extension  against numerical simulations of the disordered $\phi^4$ model and experimental data from four different ultrathin ferromagnetic films, addressing both (i) spontaneous and (ii) AC-assisted collapse of compact magnetic domains. Good agreement with experiments is obtained under conditions we discuss.

    \item \textbf{Regimes of validity}: 
    We explicitly discuss the regime of validity of the proposed geometric description, clarifying the limitations associated with finite domain-wall width, quenched disorder, and external driving.
\end{itemize}

\color{black}

The rest of the paper is organized as follows. In Sec.~\ref{sec:phi4model}, we describe elastic loops in the paradigmatic $\phi^4$ model and the numerical methods used to simulate and characterize their dynamics. In Sec.~\ref{sec:theoryvssimulations}, we present the analytical predictions and compare them with numerical simulations of the $\phi^4$ model, first addressing the linear arc-velocity response in Sec.~\ref{sec:linearhomogeneous}, followed by the nonlinear case in Sec.~\ref{sec:nonlinearhomogeneous}.  
In Sec.~\ref{sec:theoryvsexperiments}, we test selected theoretical predictions against experiments in ultrathin ferromagnetic films. In Sec.~\ref{sec:alternatingfieldexperiments}, we analyze the effect of an alternating field acting on a single domain, while in Sec.~\ref{sec:lifetimesestimates} we estimate spontaneous collapse lifetimes for domains nucleated in specific magnetic materials. We conclude the paper with a discussion and perspectives in Sec.~\ref{sec:conclusions}.

\section{Results: Theory vs numerical simulations}
\label{sec:theoryvssimulations}

In this section, we derive various analytical results for the ideal elastic loop dynamics and compare them with numerical simulations of the emerging DW loop dynamics in the $\phi^4$ model.

\subsection{Linear homogeneous arc-velocity response}
\label{sec:linearhomogeneous}

We begin by discussing the linear case \( V(x) = x/\eta \), which is expected to correspond to the homogeneous \(\phi^4\) model at \( T = 0 \) and \( r(x, y) = 0 \). In this case, the arc velocity is expected to follow the relation \( \eta v_s = \sigma \kappa_s + f \).

\subsubsection{Spontaneous collapse of a loop}
\label{sec:Spontaneouscollapseofaloop}
The simplest case is the spontaneous collapse of a compact but arbitrary shape domain due to curvature, corresponding to \( f = 0 \).  
From Eq.~(\ref{eq:arcdynamics}) and using \( v_s = \sigma \kappa_s/\eta \), we obtain, using the topological invariant of Eq. \eqref{eq:topologicalinvariant} the following equations 
\begin{align}
\frac{dA}{dt}&=-\frac{2\pi \sigma}{\eta} 
\label{eq:singleAdecay}
\\
\frac{dP}{dt}&=
-\int_{\Gamma_t}ds\; \frac{\sigma}{\eta} \kappa_s^2. 
\label{eq:singlePdecay}
\end{align}
While the area evolution is universal, the perimeter evolution depends on the initial condition, \( \Gamma_{t=0} \), and its corresponding initial curvature field \( \kappa_s = \kappa_s(t=0) \). In general, we can only assert that \( dP/dt < 0 \), and that at long times, as the domain approaches a circular shape, the evolution becomes simple and is controlled by its radius.  
Then, we have an \textit{exact} result for the area, valid for all times before the collapse time \( t < \tau \):
\begin{align}
A(t)&=A(0)-\frac{2\pi\sigma}{\eta} t, 
\label{eq:Adecay}
\end{align}
and another result for the perimeter,
\begin{align}
P(t)&\approx 
\sqrt{P^2(t^*) - 
8\pi^2 (\sigma/\eta) (t-t^*)},
\label{eq:Pdecay}
\end{align}
which is approximately valid for \( t^* < t < \tau \), in the late stages before collapse,
with $t^*$ a long enough time.
In contrast to the result for the area, Eq.~\eqref{eq:Adecay}, the perimeter result, Eq.~\eqref{eq:Pdecay}, depends on the initial shape through \( t^* \).

The lifetime, or collapse time, of the single domain is
\begin{align}
\tau= \frac{A(0)\eta}{2\pi \sigma},  
\label{eq:lifetimesingleloop}
\end{align}
for an initial simple loop enclosing an area \( A(0) \) with \textit{any} shape, thereby generalizing the result for the circle, Eq.~\eqref{eq:taucircle}. Since the perimeter also vanishes at the same time, \( P(\tau) = 0 \), we obtain  
\begin{equation}
P^2(t^*) \approx 8\pi^2 \left( \frac{\sigma}{\eta} \right)(\tau - t^*),    
\label{eq:tstarcriterion}
\end{equation}
which can be used as a practical criterion for estimating \( t^* \) from the evolution of \( P(t) \).  \color{myblue}
In the numerical comparisons, $t^*$
was taken as the earliest time at which Eq. \eqref{eq:tstarcriterion} is approximately satisfied. Beyond this point the loop remains nearly circular, and Eq. (11) accurately describes the subsequent perimeter evolution.
\color{black}

In terms of \( \tau \), we thus get
\begin{align}
    \frac{A(t)}{A(0)}&=1-\frac{t}{\tau} 
\label{eq:Asingleloop}
    \\
     \frac{P(t)}{P(t^*)}&\approx 
     \sqrt{\frac{\tau-t}{\tau-t^*}}
    \label{eq:Psingleloop}
\end{align}
with $\tau > t^* \geq 0$ in general, and $t^*=0$ holding exactly only for the circle 
\color{myblue}
(see Eq. \eqref{eq:taucircle} in Appendix \ref{circular-results}).
\color{black}

Comparing the evolution of area and perimeter in the present simple case—with non-ideal, thick interfaces obtained by solving the \(\phi^4\) model of Eq.~\eqref{eq:phi4}—to that predicted for idealized, infinitely thin interfaces [Eqs.~\eqref{eq:lifetimesingleloop}, \eqref{eq:Asingleloop}, and \eqref{eq:Psingleloop}] provides a useful sanity check.  
This comparison tests not only the modeling assumptions that link \(\phi^4\) domain walls to ideal mathematical interfaces, but also helps identify subtle numerical artifacts in the integration scheme—such as discretization or anisotropic effects—that frequently arise in standard finite-difference implementations of phase-field models.  
Controlling such artifacts is crucial, for example, when assessing the universality class of interface depinning transitions \cite{kolton2023}, or more generally, the universality class of interface growth processes \cite{BarabasiBook}.
\begin{figure}[h]
  \centering
    \includegraphics[scale=0.64]{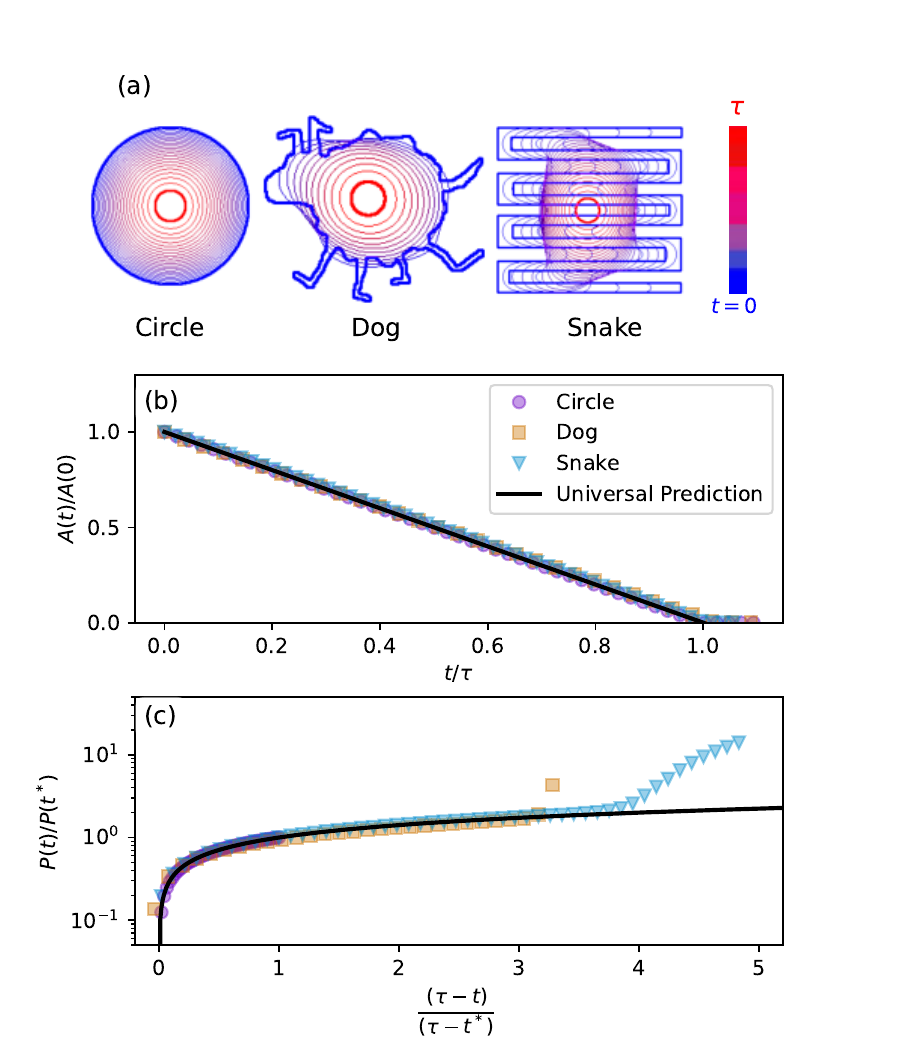}
    \caption{
    Spontaneous curvature-driven collapse of compact magnetic domains with three arbitrary initial shapes. 
    (a) Domain wall contours (loops) as a function of time. The initial configuration is shown in blue; the color bar indicates time progression. 
    (b) Normalized area, $A(t)/A(0)$, as a function of scaled time $t/\tau$, where $\tau$ is the domain lifetime given by Eq.~\eqref{eq:lifetimesingleloop}. Symbols correspond to numerical data, and the solid line is the universal prediction from Eq.~\eqref{eq:Asingleloop}. 
    (c) Normalized perimeter, $P(t)/P(t^*)$, 
    \color{myblue}
    as a function of $(\tau - t)/(\tau-t^*)$, 
    \color{black}
    where $t^*$ is a shape-dependent time at which the loop becomes nearly circular, \color{myblue} 
    estimated using Eq. \eqref{eq:tstarcriterion}\color{black}. While the perimeter evolution is not universal for $t < t^*$, it closely follows the circular collapse law, Eq.~\eqref{eq:Pdecay}, for $t > t^*$, as shown by the black solid line. 
    } 
    \label{fig:Colapso-H=0}
\end{figure}

In Fig.~\ref{fig:Colapso-H=0}(a), the evolution of domains initially shaped as a “Circle,” “Dog,” and “Snake” is shown, illustrating arbitrary examples of simply connected domains.  
In all cases, we ensured only that opposing domain walls do not come closer than the domain wall width \( \delta \), in order to avoid their attractive interactions (which are discussed later).
We clearly observe that the domains become convex and collapse into a circular shape before disappearing, as predicted for ideal, mathematically thin interfaces.  
It is worth noting that although the loops never cross themselves at a given time, they can intersect earlier configurations due to the coexistence of regions with negative and positive local curvature.

In Fig.~\ref{fig:Colapso-H=0}(b), we confirm that the decay of the area is monotonic, linear, and universal as a function of \( A(0) \) and \( \tau \), as predicted by Eqs.~\eqref{eq:Adecay} (black line) and \eqref{eq:lifetimesingleloop}.  
Conversely, in Fig.~\ref{fig:Colapso-H=0}(c), we observe that the decay of the perimeter is also monotonic, as expected from Eq.~\eqref{eq:singlePdecay}, but non-universal: shapes with many regions of large negative local curvature tend to shrink faster initially (e.g., in the Dog case, the legs and ears are rapidly absorbed, and in the Snake case, its length decreases quickly at early times).  
We also find that sufficiently close to collapse, the perimeter evolution approximately follows Eq.~\eqref{eq:Pdecay} (black line), which is in fact exact for a circle for any \( t^* \geq 0 \). 
This demonstrates that the Gage–Hamilton–Grayson theorem holds to high precision for domain walls in the \(\phi^4\) model.  
For \( t \sim \tau \), however, the model assumptions are expected to break down, as the radius of the collapsing domain becomes comparable to the domain wall width \( \delta \). Nonetheless, this regime generally represents a negligible fraction of \( \tau \) for most initial domain shapes.

Finally, we note that the lifetime \( \tau \) given in Eq.~\eqref{eq:lifetimesingleloop} is non-universal and, in terms of the \(\phi^4\) model parameters, is proportional to \( A(0)\gamma / 2\pi c \). This relation follows from the connection between the domain wall (DW) elastic tension \( \sigma \) and friction coefficient \( \eta \) in Eq.~\eqref{eq:lifetimesingleloop}, and the stiffness \( c \) and relaxation parameter \( \gamma \) in the \(\phi^4\) dynamics, through the approximation \( \sigma / \eta \approx c / \gamma \), using \( \sigma \approx \sqrt{c \epsilon_0} \) and \( \eta \approx \gamma / \sqrt{c/\epsilon_0} \).  
It is important to emphasize that for DWs to be well described as simple loops, the domain wall width \( \delta \approx \sqrt{c / \epsilon_0} \) must be sufficiently large to avoid numerical artifacts due to discretization~\cite{kolton2023}, yet small enough to prevent interaction between opposing DWs. See Appendix~\ref{AppDWWidthTensionMobility} for numerical tests that guide the appropriate choice of \(\phi^4\) parameters.

\subsubsection{Spontaneous collapse of multiple non-nested loops}

Let us now consider an initial condition consisting of \( N \) non-touching simple loops in the absence of external fields. A key property of their evolution is that the loops must remain non-touching at all times. In other words, the \textit{avoidance principle} \color{myblue}, that says that any simple curve can not cross itself during its evolution (see Appendix \ref{circular-results}), \color{black} also holds for multiple simple loops. This behavior arises from the curvature-driven dynamics: if two loops were to come into contact at a single point, the signed curvatures at that point would generally differ, leading to locally opposing normal velocities that separate the loops rather than merge them. This principle applies both to simply connected domains and to more complex domains containing nested loops, which we analyze later. At long times, each loop is expected to collapse and vanish individually, as they contribute elastic energy and are therefore thermodynamically unstable under curvature-driven evolution. \color{myblue} It is important to note that this reasoning applies to idealized mathematical elastic line loops and provides a good approximation for realistic domain walls only in the limit where the initial separation between the loops is much larger than the domain wall width, which controls their short-range interactions. \color{black}

Using the non-touching property and Eqs.~\eqref{eq:singleAdecay} and \eqref{eq:singlePdecay}, we can now derive a model for the time evolution of the total area \( A(t) = \sum_{n=1}^{N} A_n(t) \) up to the collapse time \( \tau \), defined by \( A(\tau) = 0 \). Since the loops do not interact, Eq.~\eqref{eq:singleAdecay} implies that each individual domain collapses independently at a time \( \tau_n = A_n(0)\eta / 2\pi\sigma \), which increases with the initial area \( A_n(0) \). Therefore, the time evolution of the total area \( A(t) \) reduces to a sum over the contributions of the loops that remain active at time \( t \), effectively tracking the number of surviving loops at each instant.

Let us first consider the case of simply connected domains, such that no loop contains another domain wall (DW) loop inside. In this case, if \( N_0 \) is the initial number of loops, the total area decreases monotonically at a constant rate \( -N_0 \cdot 2\pi\sigma/\eta \) until the first domain disappears. At that point, the rate of decay decreases to \( -(N_0 - 1) \cdot 2\pi\sigma/\eta \), and so on. As a result, \( A(t) \) exhibits a piecewise linear decay with discontinuous jumps in the derivative \( dA/dt \) at the collapse times \( \tau_n \). This can be expressed as
\begin{align}
\frac{dA}{dt} = -\alpha \sum_{n=1}^{N_0} \Theta(\tau_n - t) = -\alpha \sum_{n=1}^{N_0} \Theta(A_n(0) - \alpha t),
\end{align}
where \( \alpha \equiv 2\pi\sigma/\eta \), and \( \Theta(x) \) is the Heaviside step function.  
Accordingly, the total area at time \( t \) is given by
\begin{align}
A(t) = \sum_{n=1}^{N_0} \left[A_n(0) - \alpha t\right] \Theta\left(A_n(0) - \alpha t\right),
\end{align}
which depends entirely on the initial distribution of loop areas \( \{A_n(0)\}_{n=1}^{N_0} \).

Therefore, the total area is a piecewise linear, monotonically decreasing function of time, with discontinuous changes in slope occurring at \( N \) distinct times. These discontinuities are entirely determined by the initial areas of the domains and are independent of their initial shapes or spatial distribution. The complete extinction of all domains—corresponding to the convergence of the system toward the saturated background magnetization—is expected to occur at a time \( \tau \), given by
\begin{align}
\tau = \frac{A_{\mathrm{max}} \eta}{2\pi \sigma} = \frac{A_{\mathrm{max}}}{\alpha},
\label{eq:lifetimesingleloop2}
\end{align}
where \( A_{\mathrm{max}} = \max_n\{A_n(0)\} \) is the largest initial domain area.

If the individual initial areas \( A_n(0) \) are not known, but only their probability distribution is known, and the number of domains is large (\( N_0 \gg 1 \)), we can compute the average total area as
\begin{align}
\langle A(t) \rangle 
= N_0 \int_{\alpha t}^\infty da\; p(a)\, (a - \alpha t),
\end{align}
where \( p(a) \) is the probability density function of initial domain areas. From this, the time derivative of the average area becomes
\begin{align}
\frac{d\langle A \rangle }{dt}
= -\alpha N_0 \left[ 1 - F(\alpha t) \right],
\label{eq:meanareafromrandom}
\end{align}
where \( F(A) \equiv \int_0^A p(a)\, da \) is the cumulative distribution function. The evolution of \( \langle A(t) \rangle \) can thus be determined exactly from the initial area distribution. Conversely, measuring \( \langle A(t) \rangle \) as a function of time provides direct access to the cumulative distribution \( F(A) \).


Eq. \eqref{eq:meanareafromrandom} indicates that the average area and perimeter decay rates per initial domain are slower than those of a single domain, a result that can be simply interpreted from the fact that the number of active domains decreases over time as
\begin{align}
N(t) = N_0 \left[ 1 - F(\alpha t) \right].
\label{eq:Neff}
\end{align}

A simple example of application of Eq.\eqref{eq:meanareafromrandom} is a
uniform distribution of initial areas at \( t=0 \) between \( A = A_{\min} \) and \( A = A_{\max} \). Then the cumulative distribution function is given by
\[
F(A) = \frac{A - A_{\min}}{\Delta A}, \quad \text{for } A \geq A_{\min},
\]
and \( F(A) = 0 \) for \( A < A_{\min} \), where \( \Delta A = A_{\max} - A_{\min} \). 
the average area per domain evolves as
\begin{align}
\frac{\langle A\rangle}{N_0} = 
\begin{cases}
\frac{A_{\max} + A_{\min}}{2} - \alpha t & \alpha t < A_{\min}, \\
\frac{(A_{\max} - \alpha t)^2}{2 \Delta A} & A_{\min} \leq \alpha t \leq A_{\max}, \\
0 & \alpha t > A_{\max}.
\end{cases}
\label{eq:areavstuniformdist}
\end{align}
The total area thus decreases monotonically but with a positive quadratic correction, showing that the average area per domain, \( \langle A\rangle/N_0 \), decays slower than that of a single domain
. This behavior results from the time-dependent number of active domains, \( N(t) \), as given by Eq.~\eqref{eq:Neff}. All domains are certainly extinct at time \( t = \tau = A_{\max}/\alpha \), as seen from Eq.~\eqref{eq:areavstuniformdist}, since initially \( \langle A(0) \rangle/N_0 = (A_{\max} + A_{\min})/2 \). At variance with the average total area \( \langle A \rangle \), the average perimeter \( \langle P \rangle \) does not follow a universal function such as Eqs.~\eqref{eq:areavstuniformdist}, due to its dependence on the non-universal initial shapes and curvatures of the domains.
\begin{figure}[h]
  \centering
    \includegraphics[scale=0.55]{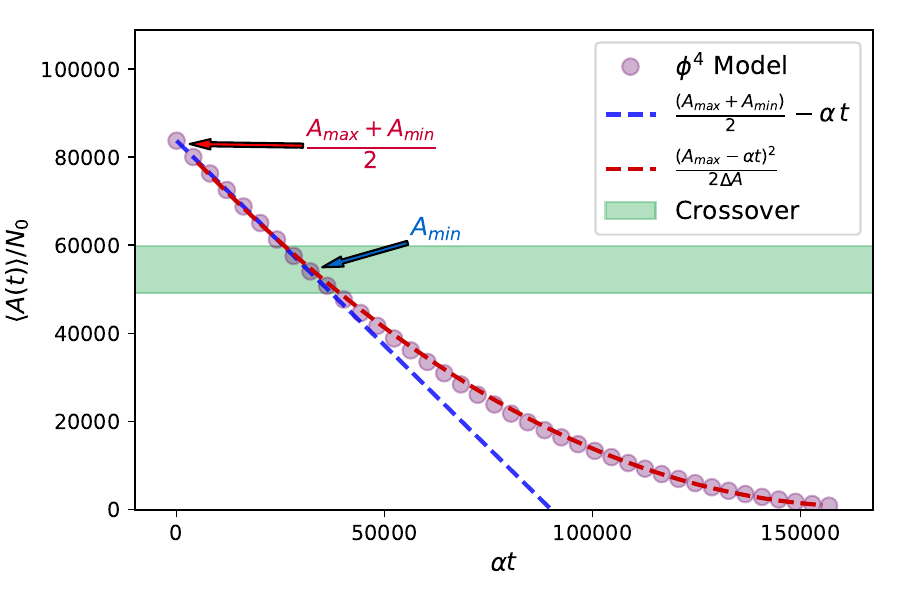}
    \caption{
Evolution of the ensemble-averaged area of an initial set of $N_0$ compact domains with a uniform random area distribution supported in $[A_{\min}, A_{\max}]$. 
The initial average area per domain, $(A_{\max} + A_{\min})/2$, decreases approximately linearly at short times, $\langle A(t) \rangle / N_0 \approx A(0)/N_0 - \alpha t$ (blue dotted line). 
A crossover (highlighted in green) occurs to a quadratic decay, $\langle A(t) \rangle / N_0 = [A_{\max} - \alpha t]^2 / 2\Delta A$ (red dotted line), where $\Delta A = A_{\max} - A_{\min}$. 
This crossover takes place at a characteristic time $t^*$ such that $\langle A(t^*) \rangle / N_0 \sim A_{\min}$, beyond which collapse events reduce the number of surviving domains.
}
\label{fig:VariosCirculos-H=0}
\end{figure}

To test the above predictions in the \( \phi^4 \) model, we exploit the independence of domains, assuming that loops do not interact at \( t=0 \) (i.e., their mutual distances are much larger than \( \delta \)). This implies that the ensemble average of a single domain's area with random initial area is \( \langle A_1 \rangle = \langle A \rangle/N_0 \), where \( A \) is the sum of the areas of the \( N_0 \) loops. Therefore, it is sufficient to perform many single-domain simulations with random initial conditions sampled from a prescribed distribution of initial areas. In Fig.~\ref{fig:VariosCirculos-H=0}, we show the numerical results for \( \langle A(t)/N_0 \rangle \) as a function of \( \alpha t \). At \( \alpha t = A_{\min} \), we observe a crossover from a linear regime to a second-order polynomial regime, in agreement with the prediction of Eq.~\eqref{eq:areavstuniformdist}. Numerical data from the \( \phi^4 \) model shows good agreement with the universal predictions derived for ideal interfaces.

\subsubsection{Spontaneous collapse of multiple loops with nesting}

Let us now consider the more general case where domains are not simply connected but can contain holes, with each hole possibly containing further holes inside, alternating the magnetization sign as one crosses them. In terms of loops, this means that the area enclosed by a loop may contain many other non-intersecting loops. In this scenario, the magnetization in the $\phi^4$ model containing $N$ domain-wall loops is given by  
\begin{align}
    M \equiv \int_{x,y} \phi = -M_{\rm s} + \phi_{\rm s} \sum_{n=1}^{N} S_n,
\end{align}
where we have defined the 
\color{myblue}
total
\color{black}
saturated background magnetization in the absence of domains as $M_{\rm s} \equiv A_{\rm sample} \phi_{\rm s}$, with $A_{\rm sample}$ the sample area. The quantity $S_n$ is the signed area enclosed by the $n$th loop,  
\begin{align}
    S_n \equiv Q_n A_n = \int_{\Gamma_n} ds \;\frac{{\bf r}_s \times {\hat t}_s}{2} \cdot {\hat z},
    \label{eq:signedarea}
\end{align}
with $A_n = |S_n|$ the absolute area of the loop and $Q_n = \pm 1$ its sign, reflecting the change in magnetization sign when crossing the domain wall from inside to outside.

In Eq.~\eqref{eq:signedarea}, the first equality defines a discrete ``charge'' $Q_n = \pm 1$ for each loop, which we choose to coincide with the sign of the magnetization enclosed by the loop. A compact domain on a negative (positive) background thus has $Q=1$ ($Q=-1$), respectively.  
In the second equality, the signed area is expressed as a line integral along the domain wall, where ${\hat t}_s$ is the tangent vector to the closed curve $\Gamma_n$ at the point ${\bf r}_s \in \Gamma_n$ (see Fig. \ref{fig:scheme}), oriented clockwise for $Q_n = -1$ and counterclockwise for $Q_n = +1$.  
Then, $M/\phi_{\rm s} + A_{\rm sample} \approx S \equiv \sum_{n=1}^N S_n$ is the total signed area, whose evolution for a fixed initial number $N$ of loops is  
\begin{align}
    S = \sum_{n=1}^{N} S_n = \sum_{n=1}^{N} Q_n \left( A_n(0) - \alpha t \right) \Theta\bigl(A_n(0) - \alpha t\bigr).
\end{align}  
Therefore, the temporal evolution of $S$ is \emph{quantized}, described by a sequence of (positive and negative) quantized slopes,  
\begin{align}
    \phi_{\rm s}^{-1} \frac{dM}{dt} = \frac{dS}{dt} = -\alpha Q,
\end{align}  
where we defined the total active charge  
\begin{align}
    Q = \sum_{\rm active} Q_n = \sum_{n=1}^N Q_n \, \Theta\bigl(A_n(0) - \alpha t \bigr),
\end{align}  
i.e., the sum over loops that have not collapsed up to time $t$. Note that at any given time $t$, $Q$ can be a positive or negative integer and necessarily $Q \to 0$ as $t \to \infty$. Moreover, for a given value of $Q$ there are many possible values of the total magnetization $M$, and hence $dS/dt = (dM/dt)/\phi_{\rm s}$ is solely controlled by $Q$. Downward/upward jumps between quantized levels of $dS/dt$ occur when a positive/negative loop collapses, respectively.

\begin{figure}[h]
  \centering
    \includegraphics[scale=0.595]{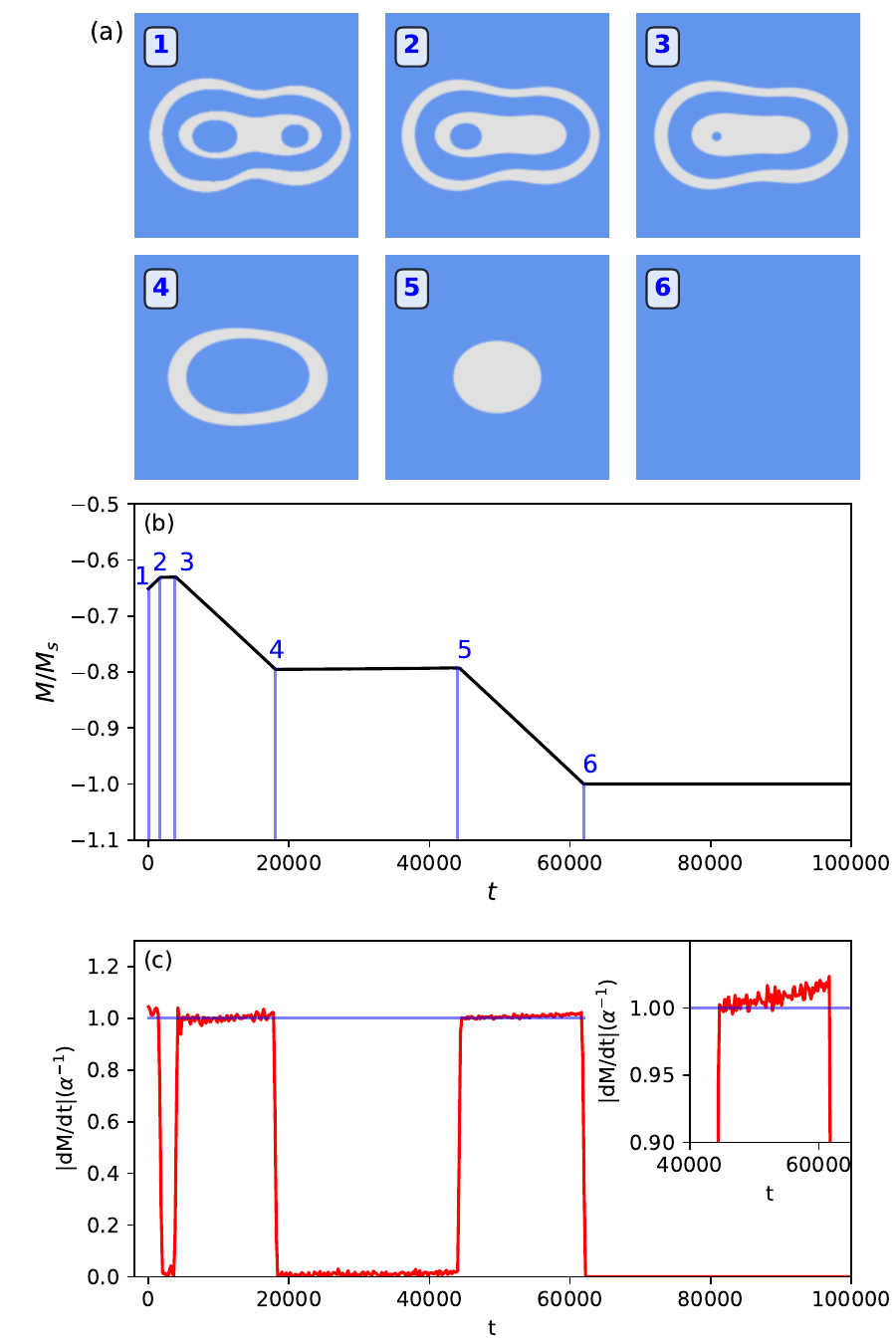}
    \caption{
\color{black}
(a) Image of local magnetization for a nested initial configuration of loops in the $\phi^4$ model until it collapses. Dark and light regions in snapshots 1–6 correspond to negatively ($\phi \approx -\phi_{\rm s}$) and positively ($\phi \approx \phi_{\rm s}$) saturated domains, respectively, at successive times. 
(b) Temporal evolution of the normalized total magnetization, 
\color{myblue}
$M/M_{\rm s}$
\color{black}
, between the snapshots shown in (a). 
(c) Normalized magnetization relaxation rate, 
\color{myblue}
$(1/\alpha M_{\rm s}) \, |dM/dt|$, 
\color{black}
over the same time interval, exhibiting discrete jumps between $n=0$ and $n=1$, associated with individual loop collapses. 
The inset shows a zoom into one of the plateaus, highlighting the acceleration induced by the attractive interaction between domain walls just prior to the final loop collapse (between snapshots 4 and 5 in panel a).
}
    \label{fig:agujeros}
\end{figure}

To illustrate the quantization of $dS/dt$ (and thus the quantization of the magnetization relaxation rate $dM/dt$), let us consider a simple example where $\alpha^{-1} dS/dt$ takes only the values $-1$, $0$, or $1$ at different times during the evolution before the collapse of all positive magnetization on a negative background.  
This can be realized by considering a ring-shaped positive magnetization domain containing two holes with negative magnetization domains of different sizes.  
In Fig.~\ref{fig:agujeros}(a) we show the evolution of this initial condition (sub-figure 1), which resembles a bold ``$\infty$'' symbol enclosed within a ring, as observed in numerical simulations of the $\phi^4$ model.  
This example highlights the different regimes of the evolution corresponding to distinct quantized values of $dS/dt$, as clearly seen in the $M$ versus $t$ curve in Fig.~\ref{fig:agujeros}(b).  
In Fig.~\ref{fig:agujeros}(c), which shows $dM/dt$ versus $t$, the jumps between quantized levels due to loop collapses are apparent.  
The inset of Fig.~\ref{fig:agujeros}(c) zooms in on one such step, revealing a small upward correction to the predicted quantized level of $dM/dt$.  
This correction is attributed to the attractive interaction, with range $\delta$, between opposite domain walls in a loop about to collapse.

\begin{figure}[h]
  \centering
    \includegraphics[scale=0.65]{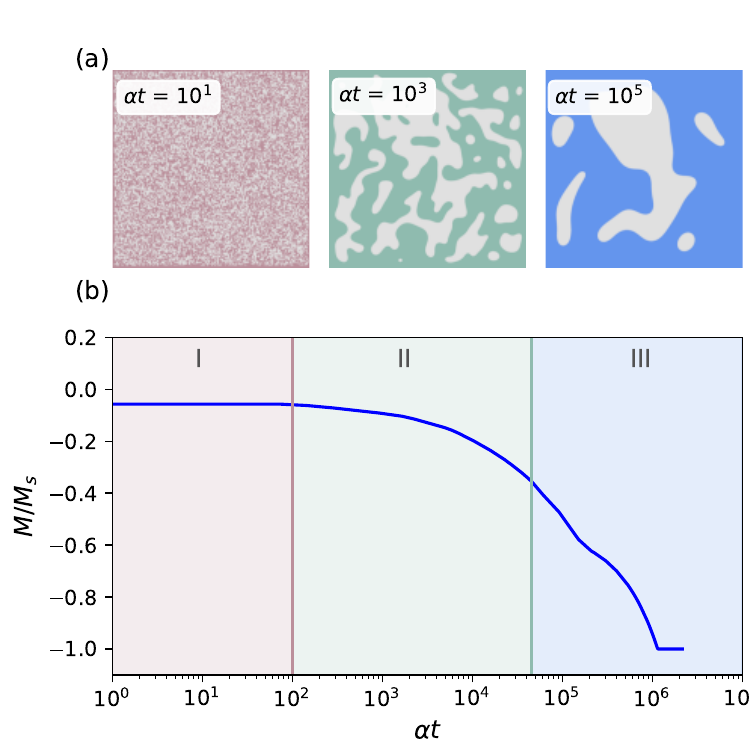}
    \caption{
Coarsening of a randomly magnetized initial configuration in the $\phi^4$ model.
(a) Snapshots of the magnetization field at different times, showing the progressive elimination of domains. 
(b) 
Normalized total magnetization $M/M_{\rm s}$ as a function of $\alpha t$, revealing three distinct dynamical regimes: 
(I) an initial smooth decay with continuous $dM/dt$, 
(II) an intermediate regime with discrete jumps of both signs associated with loop collapses and interactions, 
and (III) a late-time regime where only negative jumps remain due to the extinction of domains 
\color{myblue}
with $\phi \approx +\phi_{\rm s}$.
\color{black}
}

    \label{fig:random}
\end{figure}

It is instructive to apply the predictions for nested loops to the coarsening dynamics starting from a random initial condition, such as that shown in Fig.~\ref{fig:random}(a).  
This type of initial condition (first frame in Fig.~\ref{fig:random}(a)) is of interest as it can be experimentally or numerically realized by quenching the temperature from above to below the disorder--order transition \cite{Bray1994,Sicilia2007long}.  
For simplicity, we restrict the system to a randomly magnetized region enclosed by a negative background frame with periodic boundary conditions.  
The thin background frame is introduced to prevent the formation of percolating domains, which would require a distinct theoretical treatment.  
\begin{figure}[h]
  \centering
\includegraphics[scale=0.66]{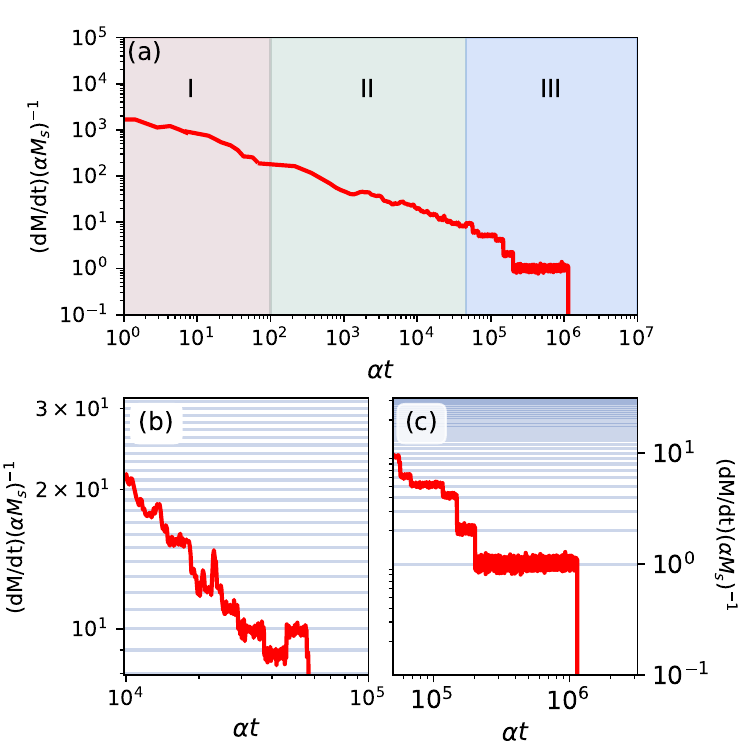}
        \caption{
Normalized magnetization relaxation rate $(\alpha M_{\rm s})^{-1} |dM/dt|$, as a function of $\alpha t$ for the coarsening dynamics shown in Fig.~\ref{fig:random}. 
(a) Full evolution, with vertical dashed lines separating the three dynamical regimes identified in Fig.~\ref{fig:random}(b). 
(b) Zoom into regime II, where both positive and negative discrete jumps in $dM/dt$ are observed. 
(c) Zoom into regime III, characterized by only negative jumps as negative domains have vanished.
Horizontal lines are guides to the eye at integer values, highlighting the quantization of $dM/dt$ in regimes II and III.
}

   \label{fig:random_Derivada}
\end{figure}
Figure~\ref{fig:random}(b) shows the evolution of the normalized magnetization, $M/M_{\rm s}$, during the coarsening process depicted in Fig.~\ref{fig:random}(a).  
While $M$ appears smooth on this scale, its time derivative, presented in Fig.~\ref{fig:random_Derivada}, exhibits numerous steps and discontinuities following an initial transient, consistent with the quantization condition $\alpha^{-1} dM/dt = n$.  
Three qualitatively distinct temporal regimes can be identified during coarsening:  
(i) At short times, $dM/dt$ varies smoothly and is not quantized. This is due partly to the incipient formation of domains, and partly to overlapping domain walls that can lead to domain coalescence or immediate collapses of small domains, violating the assumption of independent loops.  
(ii) At intermediate times, loops become decoupled, and $\alpha^{-1} dM/dt$ becomes quantized, showing both positive and negative jumps, $n \to n \pm 1$, corresponding to the collapse of domains with positive or negative magnetization (see Fig.~\ref{fig:random_Derivada}(b)).  
(iii) At long times, $\alpha^{-1} dM/dt$ remains quantized but exhibits only negative jumps (Fig.~\ref{fig:random_Derivada}(c)) since negative domains embedded in a positive background vanish before all positive domains collapse.

If the random initial condition is drawn from a distribution of signed areas, Eq.~\eqref{eq:meanareafromrandom} for the sampled average magnetization or total signed area can be generalized as  
\begin{align}
\phi_{\rm s}^{-1}\frac{d\langle M \rangle}{dt} = \frac{d\langle S \rangle}{dt} = -\alpha \sum_{Q=\pm 1} N_Q(0) \, Q \, \bigl[1 - F_Q(\alpha t)\bigr],
\label{eq:meansignedarearatefromrandom}
\end{align}
where $F_Q(a)$ and $N_Q(0)$ denote the cumulative (positive) area distribution function and the number of loops with charge $Q = \pm 1$, respectively, just after the loops decoupling transient. 
Interestingly, if $N_+(0) = N_-(0)$ and $F_+(a) = F_-(a)$, Eq.~\eqref{eq:meansignedarearatefromrandom} predicts a constant mean magnetization. However, this scenario is unphysical in our setup, since negative compact domains nucleate only inside positive domains.  
Equation~\eqref{eq:meansignedarearatefromrandom} is expected to hold after the transient regime in which loops decouple, and can thus be used to extract information about the distributions $F_Q(a)$ and numbers $N_Q(0)$ of loops after the short decoupling transient.  
Moreover, it implies that $d\langle S \rangle/dt$ is generally non-monotonic until the extinction time $t^*$ at which negative domains vanish, satisfying $F_-(\alpha t^*)=1$. For times $t > t^*$, the average signed area decreases monotonically, $d\langle S \rangle/dt < 0$.

\subsubsection{Single loop in a uniform field
}
\label{sec:singleloopuniformfield}

In the presence of a uniform field or pressure (i.e., a finite driving force \( f \)), the conclusions drawn for the spontaneous collapse of one or multiple non-intersecting simple loops must be revised. This is due to the possible emergence of singular events arising from intra- or inter-loop interactions. While such interactions are irrelevant in the \( f = 0 \) case beyond a short transient---thanks to the non-crossing theorem, which ensures that domain walls (DWs) from different loops or segments of the same loop remain sufficiently separated if they are initially non-overlapping---this is no longer true for \( f \neq 0 \). For \( f = 0 \), interactions only become relevant very close to the collapse of a loop. In contrast, for finite \( f \), interactions can influence the dynamics significantly, leading to a nontrivial evolution of the total area and perimeter, with singularities displaying well before the singularity at the collapse of individual loops. As we will show, the evolution becomes sensitive to the details of the initial condition and the nature of loop interactions.

\begin{figure}[h!]
  \centering
    \includegraphics[scale=0.56]{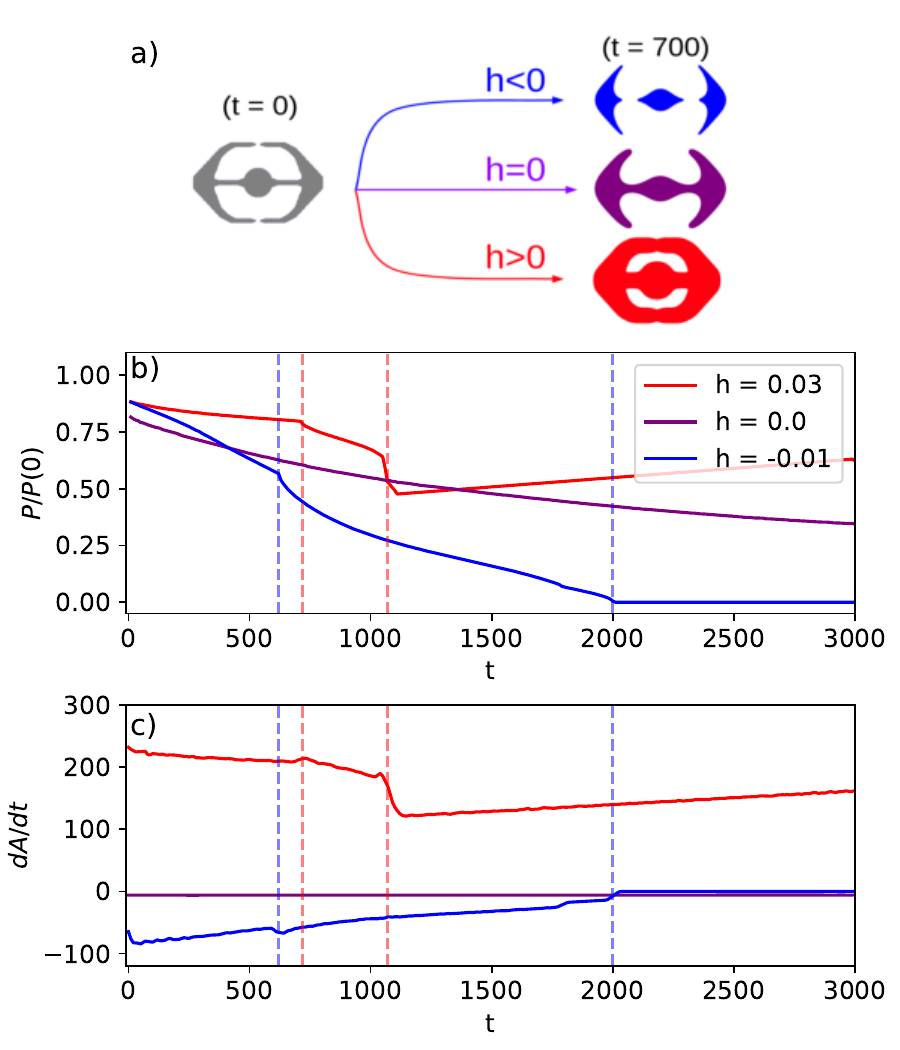}
\caption{
Illustrative example of the evolution of an eccentric domain with a spacecraft-like shape under a constant external field $h$.
(a) Starting from the initial shape at $t = 0$, the domain undergoes spontaneous collapse at zero field without changing the number of domain-wall loops.
From the same initial condition, a negative field ($h < 0$), which disfavors growth, or a positive field ($h > 0$), which promotes it, induces sparse domain-wall interaction events:
for $h < 0$, the loop splits into two loops of the same sign, while for $h > 0$ it transforms into two loops of opposite sign.
(b) Total perimeter and (c) total area as functions of time for different constant fields.
Characteristic time points are indicated:
for $h < 0$, $t_A$ denotes the splitting of the domain into three parts and $t_B$ marks the almost complete collapse of a single domain;
for $h > 0$, $t_C$ indicates the merging into a single domain containing two holes, and $t_D$ corresponds to hole closure.}
\label{fig:spaceship}
\end{figure}

Let us discuss the evolution of an arbitrary simple loop in the presence of a driving field or pressure difference \( f \). Interestingly, this general case is significantly more complex than the circular one \color{myblue}(see Eqs. \eqref{eq:circledynamicswithfield},  \eqref{eq:circleevolutionwithfield} and \eqref{eq:circlelifetimewithfield} in Appendix \ref{circular-results})\color{black}. The main reason is that the avoidance principle no longer holds: for certain initial shapes, opposing domain walls (DWs) can approach each other within a distance smaller than the wall thickness \( \delta \), leading to interaction events well before the collapse of the loop. These interactions can generate additional loops, either of the same or opposite sign compared to the initial one.

To illustrate these possibilities, in Fig.~\ref{fig:spaceship} we present simulations of the \(\phi^4\) model starting from a particular “spacecraft”-shaped initial condition. We explore the evolution under three conditions: a negative field (\(f < 0\)), a positive field (\(f > 0\)), and the unbiased case (\(f = 0\)), where \(f \propto h\).
For \(f = 0\), the area of the initial loop decreases linearly with time in a universal manner, while the perimeter evolves smoothly but non-universally (see Fig.~\ref{fig:spaceship} for \(h = 0\)). In contrast, under a negative field (\(f < 0\)), the initial loop fragments into three smaller loops of the same sign (Fig.~\ref{fig:spaceship}(a), \(h < 0\)). Under a sufficiently strong positive field (\(f > 0\)), capable of overcoming the collapse tendency due to curvature, the loop also fragments into three, but in this case the two inner loops acquire opposite signs (Fig.~\ref{fig:spaceship}(a), \(h > 0\)). In the first scenario, all resulting loops eventually collapse in finite time. In the second, the inner loops collapse while the outer loop grows indefinitely in both area and perimeter. These non-conservative events—similar in nature to the final collapse of a loop—are marked by abrupt changes in the total perimeter and area, highlighted with dashed lines in Figs.~\ref{fig:spaceship}(b) and \ref{fig:spaceship}(c). Such events are especially evident in the total perimeter evolution and become relevant whenever interacting DWs approach within a distance comparable to the wall width \( \delta \).

Although a general statement is difficult to make, between the interaction-driven, non-conserving events just described (that is, during time intervals when loops do not interact appreciably), the area $A_1$ evolution of an individual loop under an arbitrary time-dependent field \( f(t) \) satisfies
\begin{align}
    \frac{dA_1}{dt}
    = -\frac{2\pi\sigma}{\eta} + \frac{f}{\eta} P_1,
    \label{eq:relationAvsPwithf}
\end{align}
where \( P_1 \) is the perimeter of the loop. By summing over all loops, the total area \( A = \sum_{n=1}^N A_n \) and total perimeter \( P = \sum_{n=1}^N P_n \) for a given number \( N \) of loops, each with charge \( Q_n \), obey the quantized dynamical relation
\begin{align}
    \frac{dA}{dt} - \frac{f}{\eta} \sum_{n=1}^N Q_n P_n
    = -\alpha N,
    \label{eq:relationAvsPwithfwithcharge}
\end{align}
where we used that the pressure on a loop is given by \( f Q_n \). This relation effectively counts the number of active loops \( N \equiv N(t) \) of either sign at each time \( t \).
\begin{figure}
    \centering
    \includegraphics[scale=0.56]{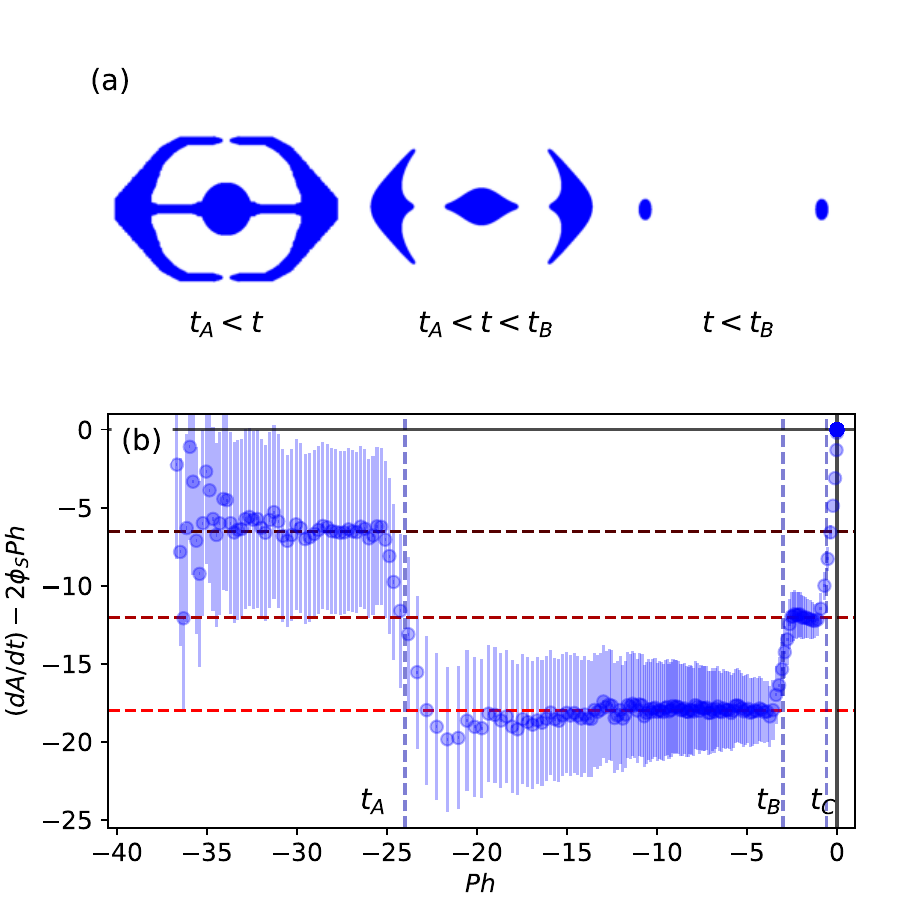}
\caption{
(a) Snapshots showing the evolution of the spacecraft-shaped domain at different time ranges. 
(b) Evolution of \( (dA/dt) - 2 \phi_{\rm s} P h \) as a function of \( P h \), using the time-dependent area \( A(t) \) and perimeter \( P(t) \) for the spacecraft shape under a constant field \( h = -0.01 \). 
Dashed lines represent fits of the form \( C N \) to the observed plateaus, which correspond— in order of decreasing \( P h \) (and increasing time)—to the number of loops \( N = 1, 3, 2, \) and \( 0 \). 
The fitted value \( C = 0.52 \pm 0.03 \) to the macroscopic geometric property is consistent with the theoretical microscopic prediction \( C \approx \sigma / (2 \phi_{\rm s}) \).
}

    \label{fig:spaceshipC}
\end{figure}

Figure~\ref{fig:spaceshipC}(b) shows that the linear relation of Eq. \eqref{eq:relationAvsPwithfwithcharge} between \( dA/dt \) and \( fP \) is well satisfied, except during the brief interaction events at \( t = t_A, t_B, t_C \), for the case of constant \( f < 0 \) in the \(\phi^4\) model, starting from the spacecraft-shaped domain shown in Fig.~\ref{fig:spaceshipC}(a). The dashed horizontal lines highlight the quantized plateaus, while the vertical dashed lines indicate interaction events: at \( t = t_A \), the initial loop splits into three \( Q = 1 \) loops; at \( t = t_B \), the central loop collapses; and at \( t = t_C \), the two remaining loops collapse nearly simultaneously due to symmetry, reaching the absorbing state \( A = 0 \), \( P = 0 \).

This type of analysis can be practical for estimating \( \sigma \) and \( \eta \) from geometric observations in real systems. In particular, the ratio between the plateau height and slope directly gives \( \sigma \), while the slope gives \( \eta \). Our simulations yield \( \sigma / 2\phi_{\rm s} \approx 0.5 \), consistent with the micromagnetic estimates \( \sigma \approx \sqrt{c \epsilon_0} \approx 1 \) and \( \phi_{\rm s} \approx 1 \) in the \(\phi^4\) model.

\subsection{Non Linear homogeneous arc-velocity response}
\label{sec:nonlinearhomogeneous}

We now consider the case where \( V \) in \eqref{eq:arcdynamics} is a \textit{non-linear} function of the local pressure. In this case, we can write, for a planar, simple, time-dependent closed curve \( \Gamma_t \) (i.e., a Jordan curve), the exact expressions:
\begin{eqnarray}
\frac{dA}{dt} &= \int_{\Gamma_t} V(f+\sigma\kappa_s) \, ds, \\
\frac{dP}{dt}&= -\int_{\Gamma_t} V(f+\sigma\kappa_s) \, \kappa_s \, ds.
\end{eqnarray}

To obtain a reduced expression in terms of \( A \) and \( P \), let us consider that \( f_d \) is the smallest characteristic field scale of \( V(f) \). If we assume that the interface is smooth enough so that the local curvature satisfies \( \sigma |\kappa| \ll f \) at every point of the interface, we can approximate
\begin{eqnarray}
V(f+\sigma \kappa) \approx
V(f) + V'(f) \sigma\kappa 
+ \frac{V''(f)}{2} \sigma^2 \kappa^2 + \dots,
\label{eq:taylorV}
\end{eqnarray}
with \( V' \) and \( V'' \) the first and second derivatives of \( V \) with respect to the field \( f \).
We can now develop perturbatively the differential equation for \( A \):
\begin{eqnarray}
\frac{dA}{dt} &\approx 
V(f) P - 2\pi \sigma V'(f) 
+ \frac{ \sigma^2 V''(f)}{2} P\langle \kappa^2 \rangle 
+ \dots
\label{eq:perturbdAdt}
\end{eqnarray}
To get this expression, we used \( P = \int_{\Gamma_t} ds \), the topological identity \( \int_{\Gamma_t} \kappa \, ds = -2\pi \), and we defined the mean squared curvature 
\[
\langle \kappa^2 \rangle \equiv \frac{1}{P} \int_{\Gamma_t} \kappa^2_s \, ds.
\]
We can proceed similarly for \( P \):
\begin{eqnarray}
\frac{dP}{dt}
\approx& 2\pi V(f) - \sigma V'(f) P \langle \kappa^2 \rangle 
- \frac{\sigma^2 V''(f)}{2} P \langle \kappa^3 \rangle 
+ \dots
\label{eq:perturbdPdt}
\end{eqnarray}

To obtain a closed approximate dynamical relation in terms of \( A \) and \( P \), we can truncate the series for \( V \) at linear order in \( \sigma \kappa \), yielding
\begin{eqnarray}
\frac{dA}{dt}&\approx& 
V(f) P - 2\pi \sigma V'(f),
\label{eq:AvstwithVoff}
\end{eqnarray}
which generalizes Eq.  \eqref{eq:relationAvsPwithf} for a known non-linear arc-velocity function \( V(f) \) for a single loop. Therefore, a fit of \( dA/dt \) versus \( V(f)P \) over a given time range is linear, 
\color{myblue}
and the intercept yields \( 2 \pi \sigma V'(f)\) and hence $\sigma$ because $V'(f)$ is known. Note that $V(f)$ and $V'(f)$, which 
\color{myblue}
represent 
\color{black}
the non-linear arc velocity response, can be known by measuring the velocity of flat in average domain walls, or segments of a large domain wall loop. 
\color{black}

Eq. \eqref{eq:relationAvsPwithf}, 
which also allows time-dependent field \( f \equiv f(t) \), 
is valid for any initial condition that is smooth enough to satisfy \( \sigma |\kappa_s| \ll f \) during a time interval that excludes the short and sparse interaction events between arcs of the same or different loops.  

At variance with the spontaneous \( f=0 \) collapse, Eqs.~\eqref{eq:AvstwithVoff} cannot be solved directly because they require knowledge of \( P \), which in turn depends on \( \langle \kappa^2 \rangle \), a quantity that cannot generally be expressed as a function of \( P \) and \( A \) alone. 

Eq.~\eqref{eq:AvstwithVoff} follows from truncating the Taylor expansion of Eq.~\eqref{eq:perturbdPdt} under the assumption $\sigma |\kappa| \ll f$. For a fixed driving field $f$, this condition, as well as the assumption of a homogeneous arc-velocity response, may be violated in the presence of strong disorder, which can induce large local curvatures.  
We note, however, that the local curvature $\kappa$ also depends on the level of coarse graining or spatial resolution used to detect the domain wall, as well as on the intrinsic domain-wall width. Since providing a quantitative assessment of these effects in all cases is nontrivial, in the following sections we instead directly confront the theoretical predictions with numerical simulations including disorder and with experimental observations.




\subsubsection{Alternating drive relations}
A typical protocol in magnetic domain-wall experiments with imaging is to apply an AC field or force $f(t)$ consisting of alternating positive and negative square pulses of identical duration \( \tau_1 \) and amplitude \( f_0 > 0 \), separated by a time interval \( \tau - 2\tau_1 \), resulting in a train of pulses with periodicity \( \tau \) (see Fig \ref{fig:Protocolo}(c) in Sec. \ref{sec:AppExperiments}). Since the spontaneous dynamics is much slower than the field-assisted one, snapshots of the domain are more conveniently taken between pulses, when \( f = 0 \). We will thus be particularly interested in describing the domain-wall loop at times corresponding to the end of each period \( \tau \) of the AC drive, as a function of the cycle number \( p \). In what follows we will consider the case $2\tau_1=\tau$ in order to neglect the $f=0$ relaxation in each cycle.

Our starting point is Eq.~\eqref{eq:AvstwithVoff} for \( dA/dt \), and our goal is to derive an approximate differential equation for the difference \( A_{p+1} - A_p \) between consecutive pulses. Integrating Eq.~\eqref{eq:AvstwithVoff} over one period around pulse \( p \), we obtain
\begin{eqnarray}
    A_{p+1}-A_p = \int_{p\tau}^{(p+1)\tau} \! du \, [V(f)P - 2\pi \sigma V'(f)].
    \label{eq:Ap+1-Ap}
\end{eqnarray}
To approximate this equation, it is useful to derive a general formula. Let us consider a function \( X(t) \) that varies very smoothly within each period \( \tau \), and a function \( g(x) \) which is either odd or even. We aim to compute the integral
\begin{equation}
    I_n \equiv \int_{n\tau}^{(n+1)\tau} g(f(t)) X(t)\,dt.
\end{equation} 
Using the trapezoidal rule, we obtain
\begin{eqnarray}
    I_n 
    &\approx& 
    g(f_0) (X_{n+1/2}+X_n) \frac{\tau}{4} \nonumber \\
    && + \, g(-f_0) (X_{n+1/2}+X_{n+1}) \frac{\tau}{4}.
\end{eqnarray}
We now consider the cases where \( g(x) \) is:
\begin{itemize}
    \item \textit{Odd}: \( g(x) = -g(-x) \), \( g'(x) = g'(-x) \);
    \item \textit{Even}: \( g(x) = g(-x) \), \( g'(x) = -g'(-x) \).
\end{itemize}
In these cases, and assuming \( X_n \) is a smooth function of \( n \), the integral simplifies to:
\begin{equation}
    I_n \approx 
    \begin{cases}
        -g(f_0) \dfrac{dX}{dn} \dfrac{\tau}{4}, & \text{if } g \text{ is odd}, \\
        g(f_0) X_n \tau, & \text{if } g \text{ is even}.
    \end{cases}
\end{equation}
Identifying \( V(f) \) as an odd function and \( V'(f) \) as an even function, we apply this approximation to Eq.~\eqref{eq:Ap+1-Ap} and obtain
\begin{eqnarray}
    \frac{dA}{dp} \approx  
    -V(f_0) \frac{dP}{dp} \frac{\tau}{4} - 2\pi \sigma V'(f_0) \tau.
\end{eqnarray}
This expression can be rewritten as
\begin{eqnarray}
    \frac{d\Lambda}{dp}
    \approx -\sigma,
    \label{eq:lambdavsp}
\end{eqnarray}
where we have introduced the geometrical observable
\begin{eqnarray}
    \Lambda \equiv \frac{A + V(f_0) P \tau/4}{2\pi V'(f_0) \tau},
    \label{eq:lambdadef}
\end{eqnarray}
which can be experimentally determined at each period \( p \).
It is worth noting that in the limit \( f_0 = 0 \), Eqs.~\eqref{eq:lambdavsp} and \eqref{eq:lambdadef} reduce to the universal formula for spontaneous collapse, Eq.~\eqref{eq:singleAdecay}, with the identification \( V'(0) = 1/\eta \) and \( V(0) = 0 \). 
It is also worth noting that in the special case where $\lim_{f\to 0} V(f),V'(f) \to 0$, Eqs.~\eqref{eq:lambdavsp} and \eqref{eq:lambdadef} also hold if $\tau > 2\tau_1$ because in Eq.\eqref{eq:Ap+1-Ap} the area can not change if $f=0$.

\color{myblue}
In the absence of external driving, the enclosed area $A$ is the natural
geometrical observable because it decreases at a universal constant rate
under curvature-driven collapse, independently of the loop shape.
Under an alternating field, however, the instantaneous area contains both
an irreversible contribution associated with the net collapse and a reversible
oscillatory ("breathing") contribution induced by the field-driven motion
of the domain wall. The combination
$A+V(f_0)P\tau/4$ removes, to leading order, this reversible contribution
and can therefore be interpreted as an effective geometric area governing
the irreversible collapse dynamics. The normalization by
$2\pi V'(f_0)\tau$ accounts for the field-dependent differential mobility
of the wall, or equivalently its effective friction, since $V'(f_0)$
generalizes the constant mobility $1/\eta$ of the linear Allen--Cahn
regime. Consequently, $\Lambda$ plays, for AC-assisted collapse with a
nonlinear velocity response, the same role that the enclosed area plays
for spontaneous curvature-driven collapse: it is a normalized effective
area whose evolution is largely independent of both the reversible
oscillatory motion induced by the alternating field and the particular
nonlinear velocity-field response.
\color{black}

In the following sections we test Eqs. \eqref{eq:lambdavsp} and \eqref{eq:lambdadef} against numerical simulations of the $\phi^4$ model without and with disorder, and also against experimental data obtained with alternating fields. 
To translate the formula to the $\phi^4$ model where $h$ is the driving field, we have to take into account that $V'(h) \equiv V'(f) \partial_h f = V'(f)2\phi_{\rm s}$ 
\color{myblue}
so ${\tilde \Lambda} = \Lambda / 2\phi_{\rm s}$. 
\color{black}
We then define \begin{eqnarray}
    \tilde{\Lambda} &\equiv \frac{A + V(h_0) P \tau/4}{2\pi V'(h_0) \tau},
\label{eq:lambdatildedef}
\end{eqnarray}
and predict, for the magnetic model 
\begin{eqnarray}
    \partial_p {\tilde \Lambda} &= -C \approx -\frac{\sigma}{2\phi_{\rm s}}.
\label{eq:lambdatildepred}
\end{eqnarray}

\subsubsection{Alternating drive simulations without disorder}
\label{sec:AlternatingDriveSimulationsWithoutDisorder}

\begin{figure}[h!]
  \centering 
    \includegraphics[scale=0.55]{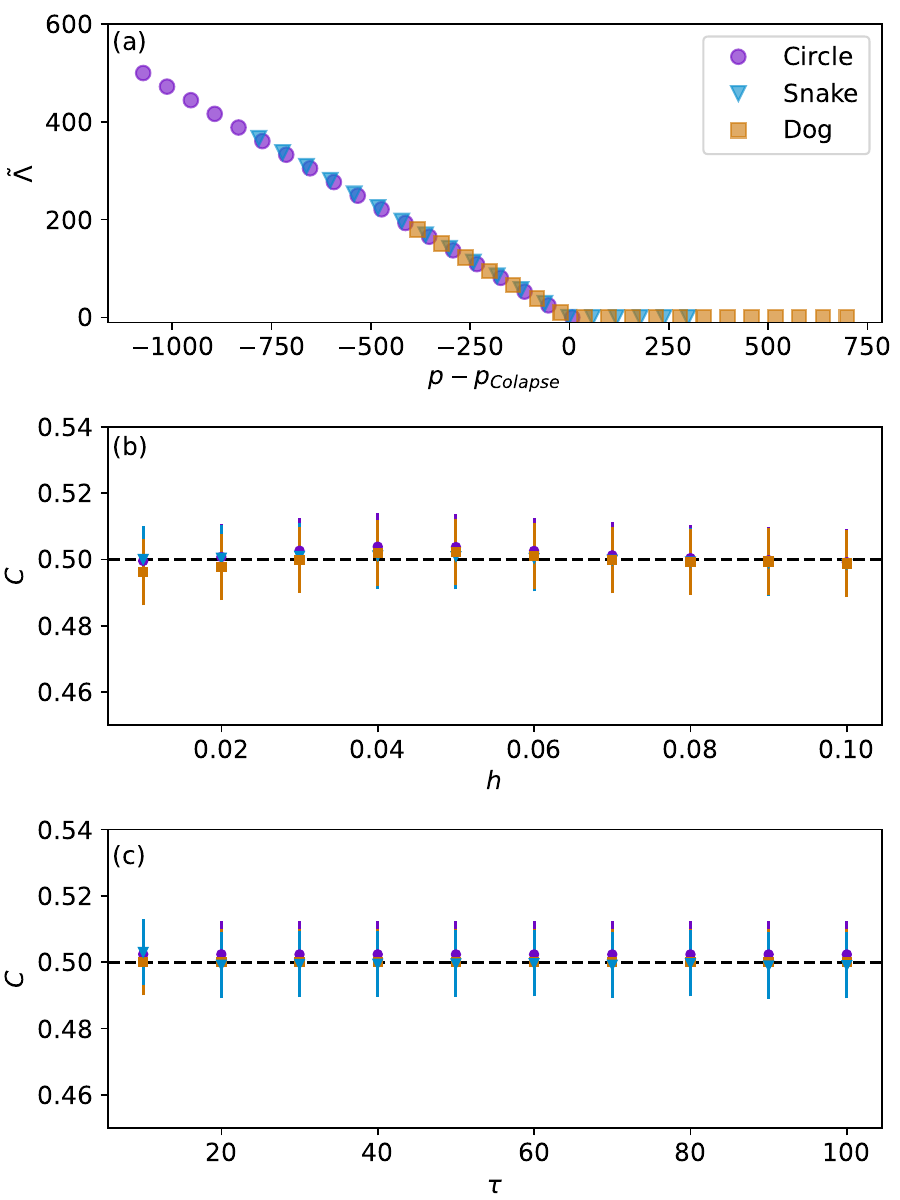}
\caption{
(a) Evolution of 
\color{myblue}
${\tilde \Lambda} = \Lambda / 2\phi_{\rm s}$ 
\color{black}
[Eq.~\eqref{eq:lambdadef}] for different initial compact domain shapes, as in Fig.~\ref{fig:Colapso-H=0}, under an alternating field with amplitude \( h = 0.1 \) and period \( \tau = 20 \). 
(b) Fitted values of \( C \) as a function of the external field amplitude \( h \), for fixed \( \tau = 20 \). 
(c) Fitted values of \( C \) as a function of the period \( \tau \), for fixed \( h = 0.1 \).
}
    \label{fig:phi4acclean}
\end{figure}

In Fig.~\ref{fig:phi4acclean}, we present results from numerical simulations of the $\phi^4$ model without disorder, using the same initial shapes as in Fig.~\ref{fig:Colapso-H=0}, but now subjected to an alternating square-wave field \( h(t) \). Each cycle consists of a positive square pulse of amplitude \( h \) and duration \( \tau_1 = \tau/2 \), immediately followed by a negative pulse of the same amplitude and duration. We investigate the response to this alternating field for various values of the period \( \tau \) and amplitude \( h \), ensuring that \( h \) is small enough to avoid coalescence or splitting of the initial loops.

Figure~\ref{fig:phi4acclean}(a) shows that the observable 
\color{myblue}
\( \tilde{\Lambda} \equiv \Lambda /  2\phi_{\rm s} \) 
\color{black}
[see Eq.~\eqref{eq:lambdadef}] closely follows the universal prediction of Eq.~\eqref{eq:lambdavsp} as a function of the number of field cycles, up to the point of collapse for the different initial shapes. From the derivative \( \partial_p \tilde{\Lambda} = -C \), we extract the parameter $C$.

In Fig.~\ref{fig:phi4acclean}(b), we show that the fitted values of \( C \) are in good agreement with the theoretical prediction 
$C = \sigma/2\phi_{\rm s} \approx 1/2$
\color{myblue}
for the specific values of $\sigma=1$ and $\phi_s=1$
we are using,
\color{black}
across a broad range of pulse counts. 
\color{myblue}
Figure~\ref{fig:phi4acclean}(b) 
\color{black}
demonstrates that these values remain robust over a wide range of pulse amplitudes \( h \), within statistical error bars. Likewise,
\color{myblue}
Fig.~\ref{fig:phi4acclean}(c) 
\color{black}
shows that \( C \) remains consistent under variations in the pulse duration \( \tau \).

It is important to stress that values of \( h \) and \( \tau \) exceeding certain shape-dependent thresholds may compromise the integrity of the initial domain, as discussed in Sec.~\ref{sec:singleloopuniformfield}. If the initial domain remains otherwise compact, these simulations illustrate a practical method for extracting microscopic parameters from dynamical geometric observables involving only \( A \) and \( P \), along with the DC velocity-field characteristic \( V(f) \).

\subsubsection{Alternating drive simulations with disorder}
\label{sec:AlternatingDriveSimulationsWithDisorder}

\begin{figure}[h!]
  \centering
\includegraphics[width=\linewidth]{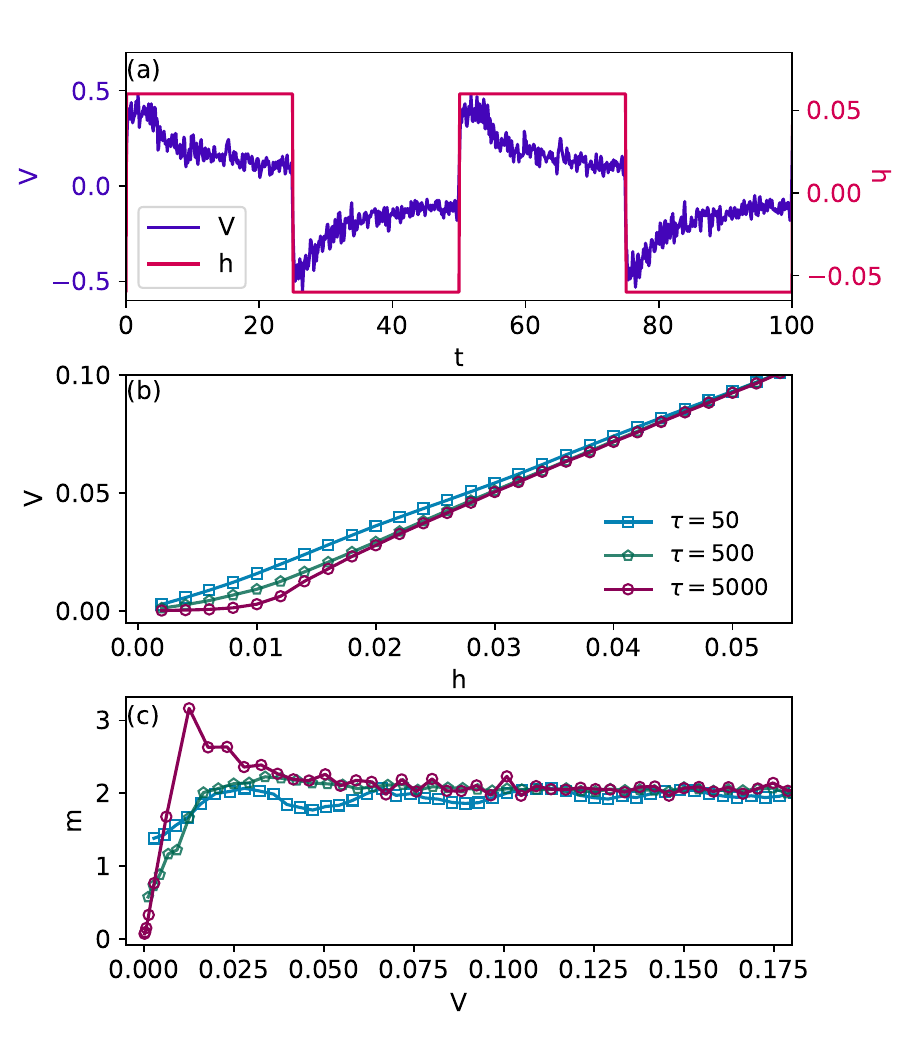}
    \caption{(a) Instantaneous domain-wall velocity $v(t)$ of a flat domain wall under a square-pulse field (blue line, left axis), together with the applied field (fuchsia line, right axis), for a simulation with disorder strength $r_0 = 1.0$, ac amplitude $h = 0.05$, and ac period $\tau = 50$.  
(b) Average ac velocity $V = \langle |v| \rangle$ as a function of the applied field $h$ \color{myblue} for $r_0=0.3$\color{black}.  
(c) Average ac mobility $m = dV/dh$ \color{myblue} obtained from (b),\color{black} as a function of $V$ for different values of $\tau$. In the limit $\tau \to \infty$, a depinning transition is observed (see Appendix).
}
    \label{fig:phi4acdisorder}
\end{figure}

\begin{figure}[h!]
  \centering    
  \includegraphics[width=\linewidth]{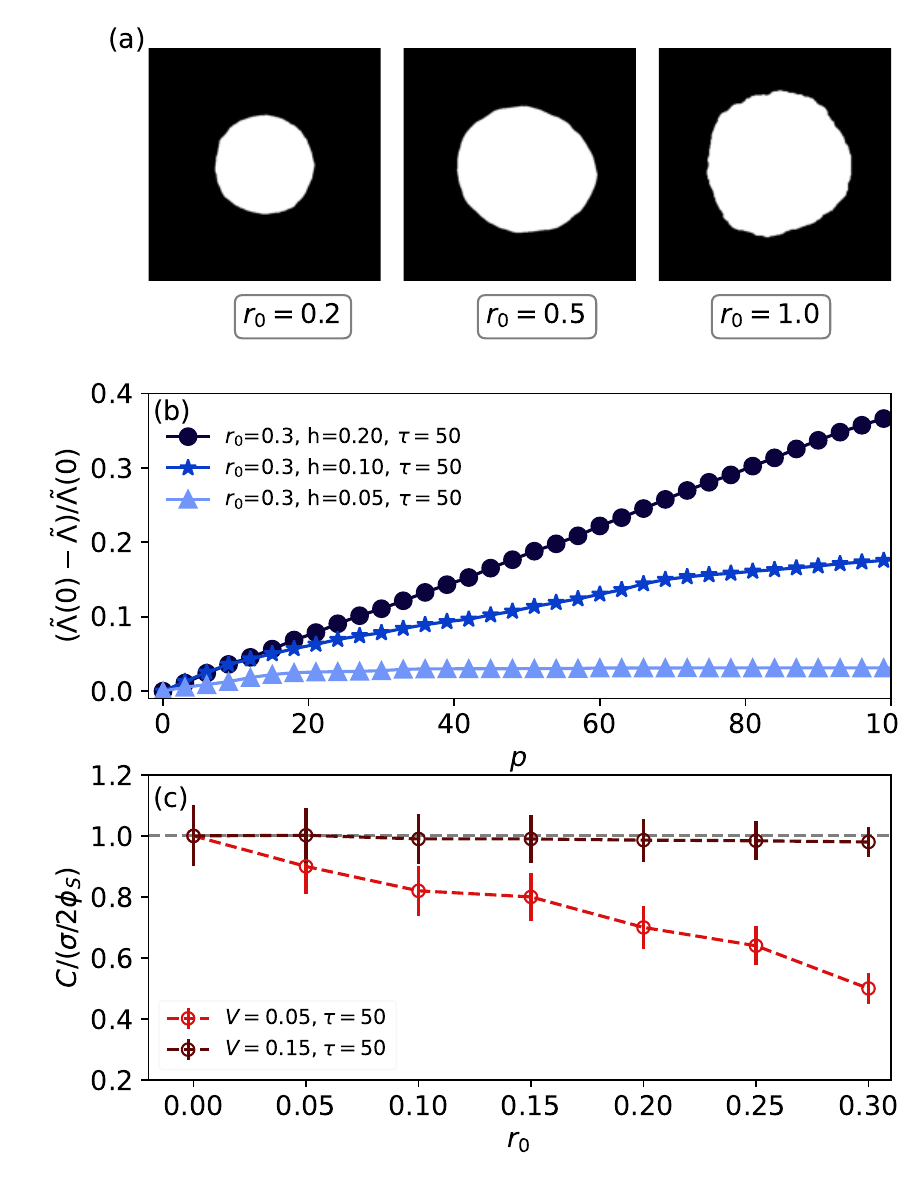}
  \caption{AC-assisted collapse of $\phi^4$ domain-wall loops under a square AC field ($\tau = 50$). 
  (a) Configurations after $p = 200$ pulses for different disorder strengths ($h=0.05$). 
  (b) Effective area $(\tilde{\Lambda}(0) - \tilde{\Lambda})$ [Eq.~\eqref{eq:lambdatildedef}] vs $p$ for $r_0 = 0.3$ and AC amplitudes $h = 0.05$–$0.20$. 
  (c) Fitted $C$ from the initial slope $-\partial_p \tilde{\Lambda}$ compared with $\sigma / 2\phi_{\rm s}$, as a function of disorder strength $r_0$, for two different AC mean velocity amplitudes.}
  \label{fig:phi4acdisorder2}
\end{figure}

We now test Eqs.~\eqref{eq:lambdatildedef} and \eqref{eq:lambdatildepred} using simulations of the $\phi^4$ model with quenched disorder at zero temperature, i.e., with a non-zero $r(x,y)$ in Eq.~\eqref{eq:phi4}, under the same alternating square-pulse field $h(t)$ described in the previous section. This study is directly relevant to experiments, where some degree of disorder is typically unavoidable.

In the presence of disorder, the assumption of a homogeneous and time-independent arc-velocity response used in Eq.~\eqref{eq:arcdynamics} becomes only approximate.  Spatially varying pinning forces contribute to the total force acting along the domain-wall contour, in addition to the elastic (curvature) and external field forces. Moreover, the arc-velocity response may acquire temporal dependence, since pinning is governed by a correlation length that characterizes how well the elastic interface adapts to the underlying disorder landscape. As a result, slow relaxation and velocity overshoots following sudden changes in the ac driving force are expected. 

In general, one should expect an arc-velocity of the form $v_s = V(\sigma \kappa_s + f, \mathbf{r}_s, t)$, which in principle invalidates the assumptions underlying Eqs.~\eqref{eq:lambdavsp} and \eqref{eq:lambdadef}, where $V$ is assumed to be independent of position $\mathbf{r}_s$ and time $t$. Additionally, these equations presume that the curvature-induced effective field $C\kappa$ is small compared to the amplitude of the applied magnetic field pulse $h$, so that the truncated Taylor expansion in Eq.~\eqref{eq:taylorV} remains valid. It is therefore not a priori clear how disorder may influence the local curvature $\kappa_s$. For this reason, it is important to test these approximations across different disorder strengths and ac driving fields.

To estimate $C$ from the ac-assisted bubble collapse, we fit $\tilde \Lambda$ [Eq.~\eqref{eq:lambdatildedef}] as a function of the pulse number $p$. 
A crucial input of the model is the stationary mobility $V'(h)$ of a macroscopically flat domain wall. In the absence of disorder, we showed in Fig.~\ref{fig:phi4acclean} that excellent agreement with the prediction is obtained by using the bare $\phi^4$ mobility $V'(h)\approx 2$ (in units of $\phi_{\rm s}/\eta$), indicating that the response of the domain wall to an ac drive involves no appreciable internal relaxation and is thus equivalent to the dc case. In this regime, the velocity response coincides with the linear response of a homogeneous medium.

In contrast, Fig.~\ref{fig:phi4acdisorder}(a) shows that in the presence of disorder, illustrated here with \color{myblue}$r_0=1.0$\color{black}, the ac drive produces the anticipated velocity overshoots, followed by a decay in magnitude toward the stationary dc velocity. If the pulse duration $\tau$ is not large enough, the velocity does not fully relax before the next overshoot, so that the average velocity response of a flat-on-average domain wall is larger than in the dc case with disorder. To quantify this effect, we define $V=\langle |v| \rangle$, with $v$ the instantaneous oscillating center-of-mass velocity of the domain wall, with $\langle v \rangle=0$. In Fig.~\ref{fig:phi4acdisorder}(b) we compare $V(h)$ for ac fields of different periods $\tau$ at $r_0=0.3$. The velocity-field characteristics are nonlinear and reveal an underlying depinning transition in the dc case at $h_{\rm d} \approx 0.1$. 
\color{myblue}
\color{black}

In the dc case (see Appendix \ref{Phi4-Details}), disorder at zero temperature leads, for interfaces smaller than a characteristic overhang length \cite{kolton2023}, to an elastic depinning transition at $h_{\rm d}$, such that for $h<h_{\rm d}$ the steady-state velocity vanishes. Just above $h_{\rm d}$, the velocity is expected to increase as a power law $V \sim (h-h_{\rm d})^{\beta}$ with $0<\beta<1$ the depinning exponent. Consequently, the differential mobility $V'(h) \sim \beta (h-h_{\rm d})^{\beta-1}$ diverges at $h_{\rm d}$ or $V\to 0$. At finite $\tau$, however, this divergence is always rounded. The shorter the pulse duration $\tau$, the smoother the curve. This is clearly seen in Fig.~\ref{fig:phi4acdisorder}(b), where we plot the differential mobility $dV/dh$ for each $\tau$ as a function of $V$. This is in agreement with the predictions of smoothening of the depinning transition by ac fields \cite{glatz2003}. At large $V$ the mobility approaches the bare value $m \approx 2$, indicating that the effect of disorder is effectively washed out by sufficiently fast motion.

With these considerations, we redefine $\tilde \Lambda$ [Eq.~\eqref{eq:lambdatildedef}] 
to use, instead of the dc mobility, the ac differential mobility $m(h,\tau)$ of macroscopically flat domain walls obtained in Fig.~\ref{fig:phi4acdisorder}(c), 
\begin{eqnarray}
    \tilde{\Lambda} &\equiv \frac{A + V(h,\tau) P \tau/4}{2\pi V'(h,\tau) \tau}.
\label{eq:lambdatildedefac}
\end{eqnarray}
This reduces to Eq.~\eqref{eq:lambdatildedef} only when $V'(h,\tau) \equiv V'(h,\infty)$ and $V(h,\tau) \equiv V(h,\infty)$, a situation realized only in the absence of disorder. 
With Eq \eqref{eq:lambdatildedefac} we expect to have a better agreement with the prediction of Eq.\eqref{eq:lambdatildepred} in the presence of disorder, with overshoots and internal relaxation affecting the average ac mobility.

With the redefinition of Eq.\eqref{eq:lambdatildedefac} we can fit $\tilde \Lambda$ as a function of $p$ for circular bubbles evolving under different disorder strengths. As shown in Fig.~\ref{fig:phi4acdisorder2}(a), after $p=200$ pulses the shrinking effect for an identical ac field becomes visibly smaller and the domain-wall loop rougher as the disorder strength is increased. In Fig.~\ref{fig:phi4acdisorder2}(b) we show how $\tilde \Lambda$ evolves with $p$ for $r_0=0.3$ but ac fields of different amplitude and identical duration $\tau=50$. We observe an approximately linear increase of $\tilde \Lambda$ with $p$ only for small $p$, followed by a crossover to a slower growth. For sufficiently small ac amplitudes $h$, e.g., $h=0.05$, $\tilde \Lambda$ can even saturate, meaning that the domain-wall loop becomes pinned in a metastable configuration around which it oscillates, so that the bubble area and perimeter remain constant on average. This situation clearly violates the prediction $\partial_p \tilde \Lambda \approx C$ at large $p$ and it may be attributed to slow relaxation process not contemplated by using the ac velocity response of flat DWs in presence of disorder.

In order to identify the conditions under which $C$ can be reliably extracted from $\tilde \Lambda$
to estimate $\sigma/2\phi_{\rm s}$
we fit its slope at small $p$ for different mean domain-wall velocities. Figure~\ref{fig:phi4acdisorder2}(c) shows that the empirical values of $C$ agree well with the prediction $C \approx \sigma/2\phi_{\rm s}$ only when the mean absolute velocity under the ac drive is sufficiently large. This provides a practical method to estimate $\sigma/2\phi_{\rm s}$ from ac experiments even in the presence of disorder, as we previously demonstrated in Fig.~\ref{fig:phi4acclean} in the clean case.


In summary, we find that in the presence of disorder, the ac method for estimating $\sigma/2\phi_{\rm s}$ from the geometrical observable $\tilde \Lambda$ works best for small $p$ and sufficiently large mean velocities, 
provided that $V \tau \ll R$, where $R$ is the average lateral size of the domain, and provided that the flat domain-wall ac responses $V(h,\tau)$ and $V'(h,\tau)$ are used. 
\color{myblue}
This requirement follows directly from the accuracy of the approximation leading to Eq. \eqref{eq:AvstwithVoff}, namely $V(f+\sigma \kappa)\approx V(f)+V'(f)\sigma \kappa$, which is valid when the curvature pressure remains small compared to the driving pressure,
$\sigma |\kappa| \ll f$. Equivalently,
$V'(f)\sigma |\kappa| \ll V(f)$,
showing that the characteristic velocity induced by curvature must remain much smaller than the field-driven velocity. This criterion provides a quantitative estimate of the regime where the observable $\Lambda$ is expected to accurately describe the collapse dynamics.
Then, 
\color{black}
the larger the mean velocity or the weaker the disorder, the wider the range of $p$ over which the fit can reliably estimate $\sigma/2\phi_{\rm s}$. Since the effect of disorder is less significant at large velocities, this result supports the validity of the uniform and constant arc-velocity approximation underlying the predictions of Eqs.~\eqref{eq:lambdatildedefac} and \eqref{eq:lambdatildepred}.

\section{Results: Theory vs Experiments}
\label{sec:theoryvsexperiments}

\begin{figure}[h!]
  \centering
    \includegraphics[width=\linewidth]{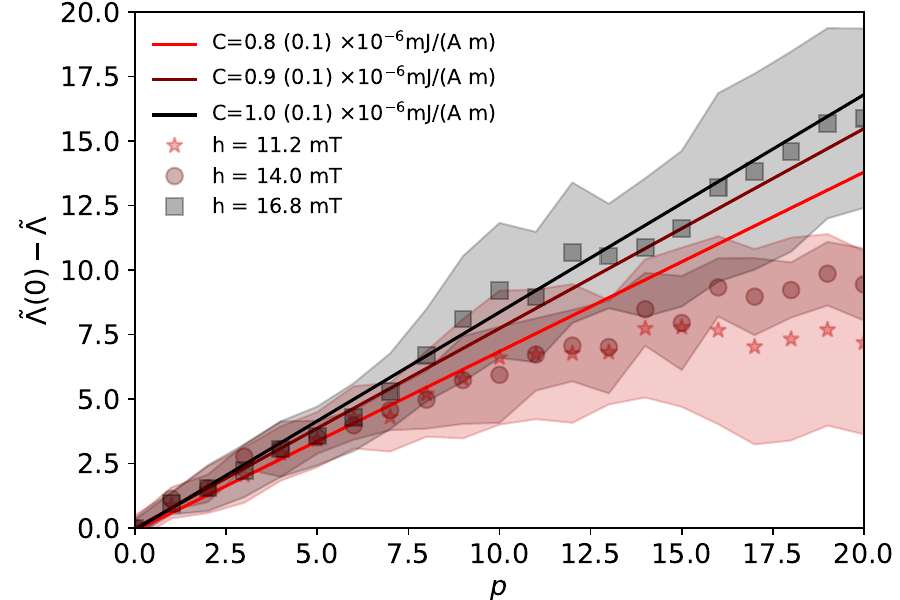}
    \caption{Evolution of $\tilde{\Lambda} - \tilde{\Lambda}(0)$ from experimental measurements on an ultrathin ferromagnetic Ta/Pt/Co/Ir/Ta film, with $\tau = 50 \, \mathrm{ms}$ and different values of $h$. Points indicate mean values, and shaded areas represent the corresponding dispersion. 
    Lines are fits $-\partial_p \tilde \Lambda \approx C$ [Eq.\eqref{eq:lambdatildepred}] for small $p$.
    }
    \label{fig:acexperiments}
\end{figure}

We now compare some of the predicted dynamical geometrical relations with single domain dynamic experiments in two dimensional ferromagnets,  
where both quenched disorder and thermal fluctuations are present. We first describe alternating field experiments with bubble domains used to estimate micromagnetic parameters of a particular ferromagnetic film, and then make estimates for the spontaneous collapse of micrometer sized domains for different films.

\subsection{Alternating field experiments in thin-film ferromagnets}
\label{sec:alternatingfieldexperiments}

To illustrate the applicability of Eqs.~\eqref{eq:lambdatildedef} and \eqref{eq:lambdatildepred} to concrete experimental situations, we now compare them with experimental data obtained from imaging domains in ultrathin ferromagnetic films subjected to an alternating field, as previously reported in Refs.~\cite{Domenichini_2019,domenichini2021} (see Sec.~\ref{sec:AppExperiments} in the Appendix for a brief description).  


In Fig.~\ref{fig:acexperiments}, we compare the evolution of ${\tilde \Lambda}$ for three different ac amplitudes at a fixed period $\tau = 50$ ms. For small pulse numbers $p$, $(\tilde \Lambda(0)-\tilde \Lambda(t))/\tilde \Lambda(0)$ grows approximately linearly, before crossing over to a slower regime. For the smallest amplitude $h=11.2$ mT, we even observe saturation, indicating that the domain becomes pinned on the experimental timescale. This behavior is qualitatively similar to that seen in $\phi^4$ model simulations with disorder, as shown in Fig.~\ref{fig:phi4acdisorder2}(b).

To avoid the saturation effect observed in Fig.~\ref{fig:acexperiments} at low AC amplitudes $h$, which violates the assumptions underlying Eq.~\eqref{eq:lambdatildepred}, we adopt the same procedure as in the simulations: we fit the slope of ${\tilde \Lambda}$ for small $p$ near $p=0$ to extract the parameter $C$. In Fig.~\ref{fig:acexperiments}, we show the fits for AC amplitudes in the range $h = 11.2$–$16.8$ mT, corresponding to mean velocities from $V = 1.5 \times 10^{-4}$ m/s up to $3.6 \times 10^{-3}$ m/s. Notably, a $\sim 50\%$ increase in $h$ produces a $\sim 2300\%$ increase in velocity, highlighting the sensitivity of the ultra-slow, thermally activated creep motion of DWs even at ambient temperature. From these fits we obtain $C \approx 0.8$–$1.0 \times 10^{-6} \, \mathrm{mJ/(A m)}$, showing the same trend observed in the zero-temperature simulations with disorder: the larger the velocity, the larger the extracted $C$. 

For our sample, the micromagnetic scale is estimated as
$\sigma = 4 \sqrt{A K_{\mathrm{eff}}} = 18 \, \mathrm{mJ/m^2}$,
with 
\color{myblue}
$\phi_{\rm s} = 1140 \, \mathrm{kA/m}$,
\color{black}
yielding
$\sigma / (2 \phi_{\rm s}) \approx 7.9 \times 10^{-6} \, \mathrm{mJ/(A\,m)}$.
This value lies within one order of magnitude of the empirical estimate
$C \approx 0.9 \times 10^{-6} \, \mathrm{mJ/(A\,m)}$, but remains smaller by a factor of approximately nine.
Thus, while the proposed method captures the correct scaling of
\color{myblue}
$\sigma / (2 \phi_{\rm s})$ 
\color{black}
in the creep regime of disordered domain walls,
a quantitative discrepancy persists.
Given that both $\sigma$ and 
\color{myblue}
$\phi_{\rm s}$ 
\color{black}
are only approximately known, and in light of the results obtained from the disordered $\phi^4$ model, we may attribute this discrepancy to the relatively small AC field amplitudes explored here, and anticipate improved agreement at larger driving fields.

Regardless of the field amplitude, we observe a systematic deviation between the linear prediction
$\tilde \Lambda(0)-\tilde \Lambda \propto p$ and the experimental data in Fig.~\ref{fig:acexperiments}
at large pulse numbers $p$.
This behavior can be plausibly associated with enhanced dynamic roughening under AC driving,
which repeatedly forces the domain wall to explore similar regions of the same disorder landscape.
More generally, it is known that fluctuating interfaces subjected to time-correlated noise exhibit
excess roughening compared to purely thermal noise, see Ref.~\cite{BarabasiBook}. Since disorder effects are
weaker at larger driving fields, this mechanism naturally explains why the deviation from linearity
sets in at larger pulse numbers for higher fields, and why after many pulses the domain wall appears
more rough than just after nucleation, as observed in Ref.~\cite{Domenichini_2019}. As a consequence, the assumption of a homogeneous arc-velocity response progressively breaks
down after a finite number of pulses. For larger AC field amplitudes, the effect of disorder during
each cycle is weaker, so that a larger number of pulses is required for this deviation to become
observable, in agreement with the experimental trends.

\subsection{Life-time of domains in ultrathin ferromagnetic films}
\label{sec:lifetimesestimates}

\begin{figure}
\begin{center}
\includegraphics[scale=0.55]{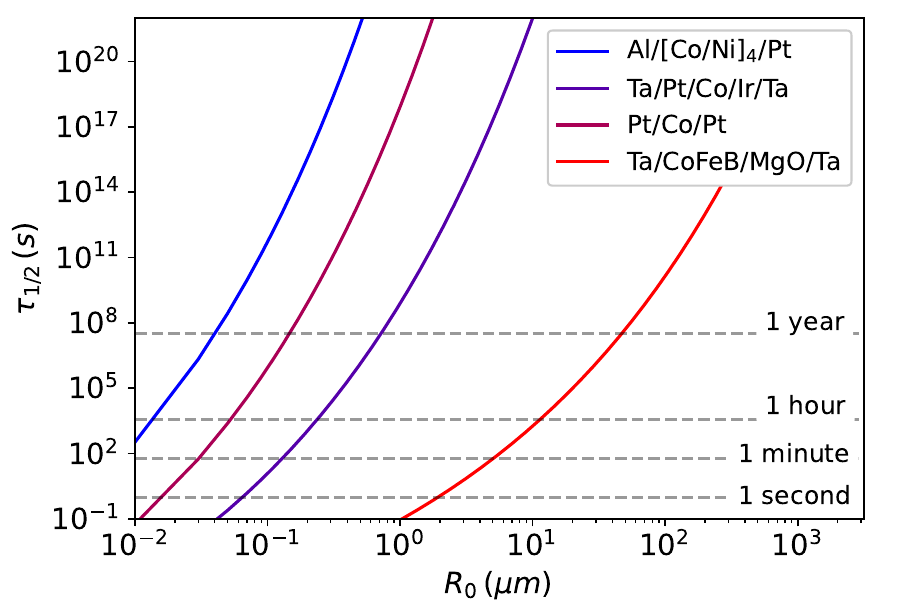}
\end{center}
\caption{
Estimated lifetimes ($\tau_{1/2}$) of a magnetic bubble as a function of its initial radius $R_0$ in the absence of applied fields, using data from ultrathin ferromagnetic films of Al/[Co/Ni]$_4$/Pt and Pt/Co/Pt  \cite{Domenichini_2019},  Ta/Pt/Co/Ir/Ta \cite{domenichini2024},
and  Ta/CoFeB/MgO/Ta 
\cite{Zhang2018}.
}
\label{fig:lifetime}
\end{figure}

\label{sec:stabilitycircular}
With our phenomenological model it is interesting to 
make rough estimates of the time needed for a circular ferromagnetic domain of initial radius 
$R_0$ to collapse spontaneously, i.e. at external magnetic field $H=0$, only acted 
by the Young-Laplace force at 
ambient temperature.
\color{myblue}
Assuming a roughly circular domain 
of radius $R$, 
$v_s \approx -dR/dt$, 
$\kappa_s \approx -1/R$, and then 
we get, from Eq. \eqref{eq:arcdynamics}, the simple closed equation
\color{black}
\begin{equation}
\frac{dR}{dt} \approx V_T(H-C/R) = -V_T(C/R)
\label{eq:circlecollapsecreep}
\end{equation}
for the time-dependent radius $R$, where 
\color{myblue}
$C=\sigma/2\phi_{\rm s}$
\color{black}
, and where $V_T(C/R_t)$ is the domain-wall arc-velocity field characteristics, 
\color{myblue}
assumed to be symmetric $V_T(C/R_t)= -
V_T(-C/R_t)$. \color{black}  
In terms of elastic loop model of Eq.~ \eqref{eq:arcdynamics}, we are using that 
\color{myblue}
$V_T(H) \equiv V(2 \phi_{\rm s} H)$ 
\color{black}
, including temperature $T$ as an extra relevant parameter, and that 
\color{myblue}
$f-\sigma/R \equiv 2 \phi_{\rm s}(H-C/R)$. 
\color{black}
It is well known that the velocity-field characteristics $V_T(H)$ can be a linear or non-linear function depending on the magnitude of $C/R$, the strength of the disorder and the temperature $T$. Two cases are worth comparing, the one without pinning, corresponding to linear velocity response associated to DW mobility, and the more realistic one with pinning, corresponding to a non-linear velocity response associated to the existence of energy barriers for DW motion. We assume that the arc-velocity response function to the local effective field (external field plus curvature pressures) is identical to the velocity-field characteristics of a flat in average domain-wall, and maintain the homogeneous response approximation to get a rough estimate.

Note that the approximation of Eq. \eqref{eq:circlecollapsecreep} for a circle is not equivalent to Eq. \eqref{eq:AvstwithVoff} for a general loop shape. Both assume homogeneous but generic \textit{nonlinear} arc-velocity response, but the former assumes a driving external field larger that the effective curvature field. Since a driving field is absent in the spontaneous collapse problem, we restrict here to circular domains, Eq. \eqref{eq:circlecollapsecreep}. Fairly circular domains are on the other hand observed in some experiments, see for instance Ref.  \cite{Zhang2018} and Appendix  \ref{sec:AppExperiments}.

\subsubsection{Without pinning}
\label{sec:stabilitycircularnopinning}
We consider first the non-disordered case, 
$V_T(H)\sim m H$, with 
\color{myblue}
$m \equiv 2\phi_{\rm s}/\eta$ 
\color{black}
the DW mobility (we are ignoring here the effects of internals degrees of freedom such at the Walker breakdown at intermediate field values). 
The equation then reduces to the Cahn-Allen equation, 
$\frac{dR}{dt} = -m \frac{C}{R}$.
The solution for this case is 
$R^2 = R^2_0 - 2 m C t$, with $R_0$ the initial radius.
Defining a ``half-radius'' lifetime $\tau_{1/2}$ 
such as $R_{\tau_{1/2}}=R_0/2$
we have $R^2_0/4 = R^2_0 - 2 m C \tau_{1/2}$ and then
\begin{equation}
\tau_{1/2}=\frac{3 R^2_0}{8 m C} \,.
\label{eq:lifetimeclean}
\end{equation}
Using the estimate 
\color{myblue}
$C = \sigma/2\phi_{\rm s} \approx 1.2 \times 10 ^{-6}\;\text{mJ/Am}$
\color{black}
\cite{domenichini2021}
and $m=430 \; \text{m/(sT)}$ \cite{Metaxas2007}
for Pt/Co/Pt films, we obtain
$\tau_{1/2}(R_0)= 7.3 \times 10^{-7} \;\text{s$ \cdot \mu$m}^{-2} \;R^2_0$.
To get an idea, a $R_0=50\;\mu \text{m}$ bubble 
has $\tau_{1/2}=1.8 \;\text{ms}$. This means that 
in the absence of disorder 
the collapse of a millimetric bubble 
should be observable in a scale of seconds.


Although this is an ideal case, it could in principle arise in weakly disordered materials with relative small domains, such that the Laplace pressure initially drives the domains walls in the fast flow regime. In other words, it could describe the last stage of a bubble collapse. 

\subsubsection{With pinning}
\label{sec:stabilitycircularpinning}

In the presence of disorder, we find that typically
$C/R_0 \ll H_{\rm d}$ for all PMOKE observable 
bubbles, typically $R_0 > 1\mu m$, where $H_{\rm d}$ is the characteristic depinning field of the sample~\cite{jeudy2018pinning}. This means that the collapse dynamics is dominated by the 
thermally activated collective creep regime~\cite{lemerle1998domain}. 
Using the circular approximation we again get a closed 
equation for $R$,
\begin{equation}
\frac{dR}{dt} = -V_T\hspace{-0.1cm}\left(\frac{C}{R}\right) 
\approx -v_0 e^{-\frac{T_d}{T}\left(\frac{R}{R_{\rm d}}\right)^{\mu}}
\label{eq:dRdtcreep}
\end{equation}
where we have defined $R_{\rm d}\equiv C/H_{\rm d}$ and used that 
$C/R_{\rm d} \gg C/R$ in order to use the universal creep law  \cite{jeudy2016universal} $V_T(H) = v_0 \exp[-(T_d/T)(H_{\rm d}/H)^\mu]$ 
with $T_d$ and $H_{\rm d}$ the characteristic temperature and depinning 
field respectively. 
Note that the creep condition is equivalent to $R \gg R_{\rm d}$, 
in consistency with the crossover from the Larkin regime 
to the random-manifold regime of the DW \cite{chauve2000creep}. 
For $R>R_{\rm d}$ the velocity force characteristics $V_T$ should include the depinning and fast flow regimes. 
The resulting equation is highly non-linear due 
to the creep response. For 
the ``half-radius life'' $\tau_{1/2}$ we obtain
\begin{equation}
\tau_{1/2}(R_0) \approx \int_{R_0/2}^{R_0} dr\; v_0^{-1} e^{\frac{T_d}{T}\left(\frac{r}{R_{\rm d}}\right)^{\mu}}.   
\end{equation}
This integral can be done analytically. 
It is however clear that since 
we are interested in 
$R_0 \gg R_{\rm d}$  and $T_d>T$
(for Pt/Co/Pt films we have for instance 
$R_{\rm d} \approx 0.02 \mu m$ and $T_d/T \approx 22.3$  
for $T\sim 293K$ and using $T_d\sim 6533\text{K}
$, see Table \ref{tab:samples} )
, the integral will be dominated by the initial radius $r\sim R_0$,
\begin{equation}
\tau_{1/2}(R_0) \approx \frac{R_0}{V_T(C/R_0)} 
\end{equation}

Therefore, while the dependence with the initial radius is quadratic 
in the absence of disorder (Eq.(\ref{eq:lifetimeclean})), 
with disorder it is a stretched-exponential. 
To get an estimate one can naively extrapolate the velocity 
field characteristics $V_T(H)$ to a very small field $H\sim C/R_0$. 
Doing so 
we obtain the 
result of Fig.\ref{fig:lifetime} 
for four different ferromagnetic films, using the values of Table \ref{tab:samples} in the Appendix, extracted from \cite{jeudy2018pinning, Domenichini_2019,domenichini2024}. 


As can be appreciated in Fig.~\ref{fig:lifetime}, micron-sized domains in typical ultrathin ferromagnetic films appear to be extremely stable, with their stability experimentally compromised only at the submicron scale.
The only exception is CoFeB, which is discussed separately in Sec.~\ref{sec:deflating}.
To understand the origin of these large differences between samples, we note that the curvature-driven creep law in Eq.~\eqref{eq:dRdtcreep} depends strongly on the ratio between the characteristic energy scale $T_d R_{\rm d}^{-\mu}$ and the thermal scale $T R^{-\mu}$.
In practice, one typically has $R > R_{\rm d}$ (otherwise the dynamics is deterministic) and $T_d > T$ at ambient temperature.
As a result, the exponential dependence of the Arrhenius activation produces differences of many orders of magnitude in the half-radius lifetime $\tau_{1/2}$, as summarized in Table~\ref{tab:samplestau}.
These lifetimes range from practically unobservable spontaneous collapses to values of the order of minutes in the case of the softer ferromagnet CoFeB.

\begin{table}[h]
\centering
\caption{
Half-radius lifetime $\tau_{1/2}$ for an initial radius $R_0 = 10\,\mu\mathrm{m}$ in different thin-film ferromagnetic samples.
}
\label{tab:samplestau}
\begin{tabular}{|l|c|c|c|c|}
\hline\hline
 & [Co/Ni]$_4$  & Pt/Co/Pt  & Pt/Co/Ir  & CoFeB \\
\hline
$\tau_{1/2}(R_0)$ (s) & $2.5 \times 10^{51}$ & $3.8 \times 10^{38}$ & $1.3 \times 10^{13}$ & $1.3 \times 10^{3}$ \\
\hline\hline
\end{tabular}
\end{table}
\color{black}

\color{black}

For the ultra stable domains in 
Al/[Co/Ni]$_4$/Pt, Pt/Co/Pt and  Ta/Pt/Co/Ir/Ta it is interesting to estimate their capacity to store bits as small domains. 
Using $1~\mathrm{cm}^2$ of a thin film to store bubble bits stable for 1 year requires radii of approximately 
$R_0 \sim 0.1~\mu\mathrm{m}$ for Pt/Co/Pt, $R_0 \sim 1.0~\mu\mathrm{m}$ for Ta/Pt/Co/Ir/Ta, and 
$R_0 \sim 3 \times10^{-2}~\mu\mathrm{m}$ for Al/[Co/Ni]$_4$/Pt, corresponding to bit densities of roughly 
$2.5\times 10^9$, $2.5\times 10^7$, and $2.8\times 10^{10}$~bits/cm$^2$, respectively. 
For comparison, current hard-disk and MRAM technologies achieve bit densities of $10^{11}$--$10^{13}$~bits/cm$^2$, 
indicating that while bubble domains can store information stably, their density is still orders of magnitude lower than modern commercial magnetic memory.

Regarding the regime of validity of the model in Eq.~\eqref{eq:dRdtcreep}, two main assumptions must be considered.
First, the radius $R$ is assumed to remain larger than $R_{\rm d}$; otherwise, the velocity response function $V_T$ must be modified to include the depinning and fast-flow regimes.
Second, even in the thermally activated regime, the domain-wall (DW) length must remain smaller than the typical size $L_{\rm opt}$ of thermal nuclei; otherwise, the universal creep law is replaced by Arrhenius activation over a finite energy barrier.
In imaging experiments, the observable radius $R$, which is limited by the magneto-optical resolution, is typically larger than both $L_{\rm opt}$ and $R_{\rm d}$.
Therefore, the use of Eq.~\eqref{eq:dRdtcreep} is justified, provided one does not attempt to model quantitatively the very final stage of the collapse, which is expected to last a time of order milliseconds (see Sec.~\ref{sec:stabilitycircularnopinning}).

\subsection{Spontaneous deflation of a magnetic bubble}
\label{sec:deflating}

As shown in Fig.~\ref{fig:lifetime}, the FeCoB samples exhibit a markedly shorter collapse lifetime than the other films. This behavior can be traced back to their comparatively low depinning temperature $T_d$ and depinning field $H_{\rm d}$, which imply a relatively large characteristic radius $R_{\rm d}$ of the order of microns. As a result, the spontaneous evolution of micrometric magnetic bubbles in these films occurs on experimentally accessible time scales—of the order of seconds—allowing their temporal dynamics to be directly observed. This extreme sensitivity of the velocity response to sample disorder is a hallmark of creep motion.

\begin{figure}[h]
  \centering
  \includegraphics[scale=0.5]{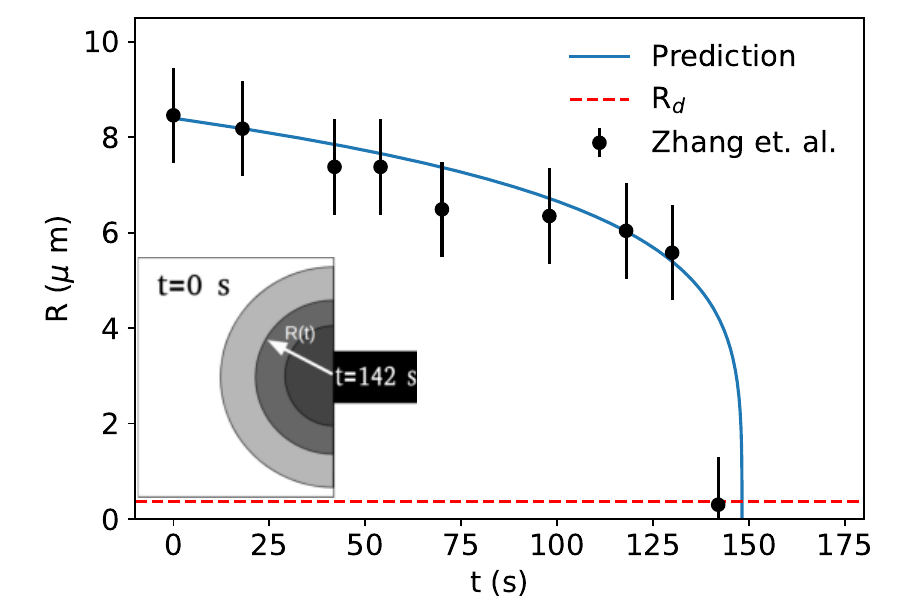}
  \caption{ 
  Radius $R$ as a function of time $t$ during the collapse of a semicircular domain, schematically shown un the inset. Black dots correspond to data digitized from Ref.~\cite{Zhang2018} for this setup, while the solid blue curve represents the theoretical prediction obtained using measured parameters from ultrathin Ta/CoFeB/MgO/Ta films~\cite{jeudy2018}. The critical radius $R_{\rm d} \simeq 0.36\,\mu\mathrm{m}$, marking the crossover from the creep to the flow regime, is indicated by the red dashed line.
  }
  \label{fig:Rd-CoFeB}
\end{figure}

The experiment reported in Ref.~\cite{Zhang2018} probes Laplace-driven collapse by first ``inflating'' an approximately semicircular magnetic domain attached to a narrow ferromagnetic strip, as illustrated in Fig.~\ref{fig:Rd-CoFeB}. In the following, we apply our geometric framework to estimate the short-time dynamics of such bubbles, using independently measured pinning and micromagnetic parameters for CoFeB samples. Figure~\ref{fig:Rd-CoFeB} shows the bubble radius as a function of time during spontaneous contraction, obtained by digitizing the data from Ref.~\cite{Zhang2018}. The initial radius is $R_0 \approx 8.5\;\mu$m.

We find that the data are reasonably well fitted by the solution of the phenomenological model given by Eq.~\eqref{eq:dRdtcreep} (solid line in Fig.~\ref{fig:Rd-CoFeB}), using the CoFeB parameters listed in Table~\ref{tab:samples}: $T_d=2230 \,\mathrm{K}$, $T=293 \,\mathrm{K}$, 
\color{myblue}
$\phi_{\rm s}=1000 \, \mathrm{kA/m}$, 
\color{black}
$H_{\rm d}=7 \, \mathrm{mT}$, $\sigma=5.1 \,\mathrm{mJ / m^{2}}$, yielding 
\color{myblue}
$R_{\rm d}=\sigma/(2\phi_{\rm s} H_{\rm d})\approx 0.36 \, \mu\mathrm{m}$. \color{black}These values are consistent, within experimental uncertainty, with those reported in Refs.~\cite{Zhang2018,jeudy2018pinning} for samples of similar composition. It is worth stressing that the same phenomenological model predicts, for the same initial radius $R_0=8.5 \, \mu$m, a practically unobservable collapse for the other, magnetically ``harder'' samples, shown in Fig.~\ref{fig:lifetime}.

The model also naturally accounts for the acceleration of the contraction observed around
$t\simeq135\,\mathrm{s}$. This apparent discontinuity originates from the extreme sensitivity of
the creep law to the ratio $R(t)/R_{\rm d}$ when $T \ll T_d$. Since the creep barriers scale as
$U(R) \sim k_B T_d [R(t)/R_{\rm d}]^{1/3}$, while the available thermal energy is fixed at
$k_B T$~\cite{ferrero2021}, the curvature-induced creep velocity increases rapidly as $R(t)$
approaches $R_{\rm d}$ from above. Indeed, the Arrhenius factor
$\exp[-U(R)/k_B T]$ decreases by about two orders of magnitude when $R$ is reduced from
$8\,\mu\mathrm{m}$ to $4\,\mu\mathrm{m}$, and by a further three orders of magnitude as $R$
approaches $R_{\rm d}$, amounting to a total change of roughly \textit{five} orders of magnitude from the
initial radius $R_0$.

We also note that when $R(t)\sim R_{\rm d}$, our description becomes increasingly approximate, as the velocity
response is expected to cross over from the creep regime to thermally rounded or fast-flow
dynamics, which would imply an even faster contraction. For sufficiently small $R$, however,
geometrical pinning effects may also become relevant for the experiment of
Ref.~\cite{Zhang2018}.

Although our results are consistent with independently measured sample parameters, within their experimental uncertainty, and with the heuristic model embodied in Eq.~\eqref{eq:dRdtcreep}, the strong sensitivity of the creep law implies that fits to the present experimental data do not allow for a precise determination of the underlying micromagnetic and pinning parameters. This limitation is intrinsic to the activated nature of creep dynamics, rather than to the geometric approach itself.

\section{Conclusions}
\label{sec:conclusions}

We have developed a geometric framework to describe the evolution of planar domain-wall loops, based on the coupled temporal evolution of their area and perimeter under both linear and nonlinear arc-velocity responses, with and without an external drive.
Analytical predictions were confronted with numerical simulations of the scalar magnetic $\phi^4$ model and with experiments in thin-film ferromagnets, allowing us to delineate the range of validity and the main limitations of the approach.
As a final application, we derived quantitative predictions for the curvature-induced spontaneous collapse of compact magnetic domains in four different samples, finding good overall agreement with experimental observations.

Beyond these specific applications, the phenomenological framework introduced here provides a practical route to inferring effective micromagnetic and pinning parameters from domain-imaging experiments in thin-film ferromagnets, and more generally offers a minimal and flexible description for the dynamics of simple elastic domain-wall loops.
\color{black}


\appendix

\section{Simple analytical predictions and known mathematical results}
\label{circular-results}

In this section we obtain simple analytical results which are useful to compare with the ones obtained and discussed for arbitrary loop shapes in the main text.

\subsection{Spontaneous collapse:}  
To illustrate the use of Eqs.~\eqref{eq:arcdynamics}, \eqref{eq:dAdt}, and \eqref{eq:dPdt}, consider the simplest case of a circular domain spontaneously collapsing due to its own curvature. Its instantaneous normal velocity response 
\color{myblue}
is given by $v_s = (\sigma / \eta) \kappa_s$,
\color{black}
corresponding to $f=0$ and $V(x) = x / \eta$ in Eq.~\eqref{eq:arcdynamics}, where $\eta$ is a friction constant. For a circle with instantaneous radius $R$, the signed curvature is simply $\kappa_s = -1/R$, and the area and perimeter are $A = \pi R^2$ and $P = 2 \pi R$, respectively. Eq.~\eqref{eq:dAdt} then implies  
\begin{align}
R = \sqrt{R_0^2 - 2 (\sigma / \eta) t},
\end{align}
where $R_0$ is the initial radius.
The lifetime of the planar bubble is then simply 
\begin{align}
\tau=R_0^2 \frac{\eta}{2\sigma} ={A_0} \frac{\eta}{2\pi \sigma}.
\label{eq:taucircle}
\end{align}
where $A_0 = \pi R_0^2$.
Remarkably however, while the second term in Eq. \eqref{eq:taucircle} is valid only for a circle, the third one, in terms of the initial area $A_0$, is  valid for \textit{any initial simple closed curve} with initial area ${A}_0$, not necessarily a circle. This follows from simply noting that for \textit{any} simple possibly evolving closed loop $\Gamma_t$, the quantity
\begin{align}
\int_{\Gamma_t} \kappa_s ds=-2\pi,     
\label{eq:topologicalinvariant}
\end{align}
is a topological invariant and a constant along the time evolution. Therefore, the total curvature is conserved and the spontaneous area collapse of a loop following Eq.\eqref{eq:arcdynamics} with $V(\sigma \kappa_s + f)=(\sigma \kappa_s + f)/\eta$ is exactly given, \color{myblue} for $f=0$, \color{black} by 
\begin{align}
\frac{dA}{dt} = \frac{\sigma}{\eta} \int_{\Gamma_t} \kappa_s ds = -2\pi \frac{\sigma}{\eta},
\label{eq:universalAdecay}
\end{align}
and then the decay $A=A_0-2\pi\sigma t/\eta$ is universal, regardless of the initial shape of area $A_0$.
The only assumptions we made are that $\Gamma_{t=0}$ is a simple smooth loop and then $\Gamma_t$ is mathematically guaranteed to remain a simple loop during its lifetime, and that the interface velocity is normal and proportional to the local curvature, which is well defined in each point of the curve for all $t$. The universal result of Eq. \eqref{eq:universalAdecay} is remarkable considering that the local signed curvature evolution obeys, in the absence of other forces other than curvature, the non-linear partial differential equation
\cite{GageHamilton1986}
\begin{align}
     \eta \partial_t \kappa_s = \sigma \partial^2_{s} \kappa_{s} + \sigma \kappa_s^3 
    \label{eq:kappaevolution}
\end{align}
which shows that $\kappa_s$ retains memory of the initial condition $\kappa_s(t=0)$ until it reaches a circular form before collapse, where $\partial^2_s \kappa=0$, 
$\kappa_s=-1/R$ and then $\dot R / R^2 = -(\sigma/\eta
)/R^3$ from Eq. \eqref{eq:kappaevolution}, as expected. 
The universal decay predicted by Eq.\eqref{eq:universalAdecay} is protected by an \textit{avoidance principle} that says that any simple curve can not cross itself during its evolution and hence it is valid until the loop disappears into a point singularity.

Although the case just discussed corresponds to an ideal, infinitely thin elastic closed interface in the particular scenario of spontaneous collapse, under certain approximations,
Eqs. (\ref{eq:dAdt}) and (\ref{eq:dPdt}) can still provide some interesting geometrical insights into the complicated dynamics of 
Eq.(\ref{eq:arcdynamics}). This applies to finite-width domain walls forming one or even multiple nested loops, each represented by a different $\Gamma_t$, as often encountered during coarsening in various phase-field or level-set models, for instance.
As an example of these insights, we show that in the case of a linear arc-velocity response $V$, the \color{myblue} total  \color{black} magnetization (or the equivalent quantity in a non-magnetic system), which is linearly related to the areas of the domains, has a time derivative that is exactly quantized in the absence of forces other than those arising from local curvatures, with discrete jumps occurring only near domain collapse events. A similar quantization property also holds for a combination of the area and the perimeter when an external field is present, with jumps induced by sparse interaction events between domain walls. Furthermore, we show that certain predictions can be extended to the general case of a non-linear arc-velocity function $V$ in Eq.~\eqref{eq:arcdynamics} by considering interfaces with weak curvature and/or sufficiently smooth arc-velocity responses. With these approximations, we derive closed dynamical relations between the area and perimeter of the loops, which prove useful for making concrete experimental predictions—such as the lifetime of domains in the absence of an external field and their evolution under an alternating drive. To validate these predictions, we compare them with simulations of the $\phi^4$ model, which captures finite-width interfaces and their short-range interactions, as well as with experimental data obtained from imaging magnetic domain walls in ultrathin ferromagnetic films.

\subsection{Evolution under a constant external field:}  
Before addressing the general case, let us begin with the evolution of an initially circular domain under a constant external field \( f \). This case is particularly simple because the domain remains circular throughout its evolution, with its instantaneous radius \( R(t) \) obeying the equation
\begin{align}
    \eta \frac{dR}{dt} = -\frac{\sigma}{R} + f,
    \label{eq:circledynamicswithfield}
\end{align}
which is \color{myblue} paradigmatic in classical nucleation theory~\cite{Bray1994}. \color{black} This equation has a single fixed point at \( R^* = \sigma/f \), indicating that if the initial radius \( R_0 > R^* \), the domain will grow asymptotically as \( R(t) \sim (f/\eta)\, t \). Conversely, if \( R_0 < R^* \), surface tension dominates over the driving force \( f \), and the domain shrinks, collapsing to \( R \to 0 \) in a finite time \( \tau \).

\color{myblue}
The implicit solution for the radius is
\begin{align}
\frac{f t}{\eta} = 
R - R_0 + \frac{\sigma}{f} \ln \left(
\frac{f R - \sigma}{f R_0 - \sigma}
\right),
\label{eq:circleevolutionwithfield}
\end{align}
and the collapse time \( \tau \) is obtained by setting \( R(\tau) = 0 \):
\begin{align}
\frac{f \tau}{\eta} = 
- R_0 - \frac{\sigma}{f} \ln \left(
\frac{\sigma - f R_0}{\sigma}
\right).
\label{eq:circlelifetimewithfield}
\end{align}
\color{black}
This expression shows that \( \tau \) is finite only if \( R_0 < R^* = \sigma/f \), and that it increases with increasing \( f \), as expected. In the limit \( f \to 0 \), we recover the field-free collapse time given in Eq.~\eqref{eq:taucircle}.

\section{Width, elastic tension, and mobility of $\phi^4$ domain walls}
\label{AppDWWidthTensionMobility} 

We show that the elastic tension $\sigma$ and friction coefficient $\eta$ of $\phi^4$ domain walls are related to the stiffness $c$ and the relaxational parameter $\gamma$ of the dynamics as
\[
\frac{\sigma}{\eta} \approx 1.14 \, \frac{c}{\gamma},
\]
and that the wall width scales as $\delta \approx \sqrt{c}$ for $\epsilon_0 = \gamma = 1$, provided $c \geq 1$. Deviations for $c < 1$ \color{myblue} arise from numerical discretization effects \cite{kolton2023}.\color{black}  

\begin{figure}[h]
  \centering
  \includegraphics[scale=0.44]{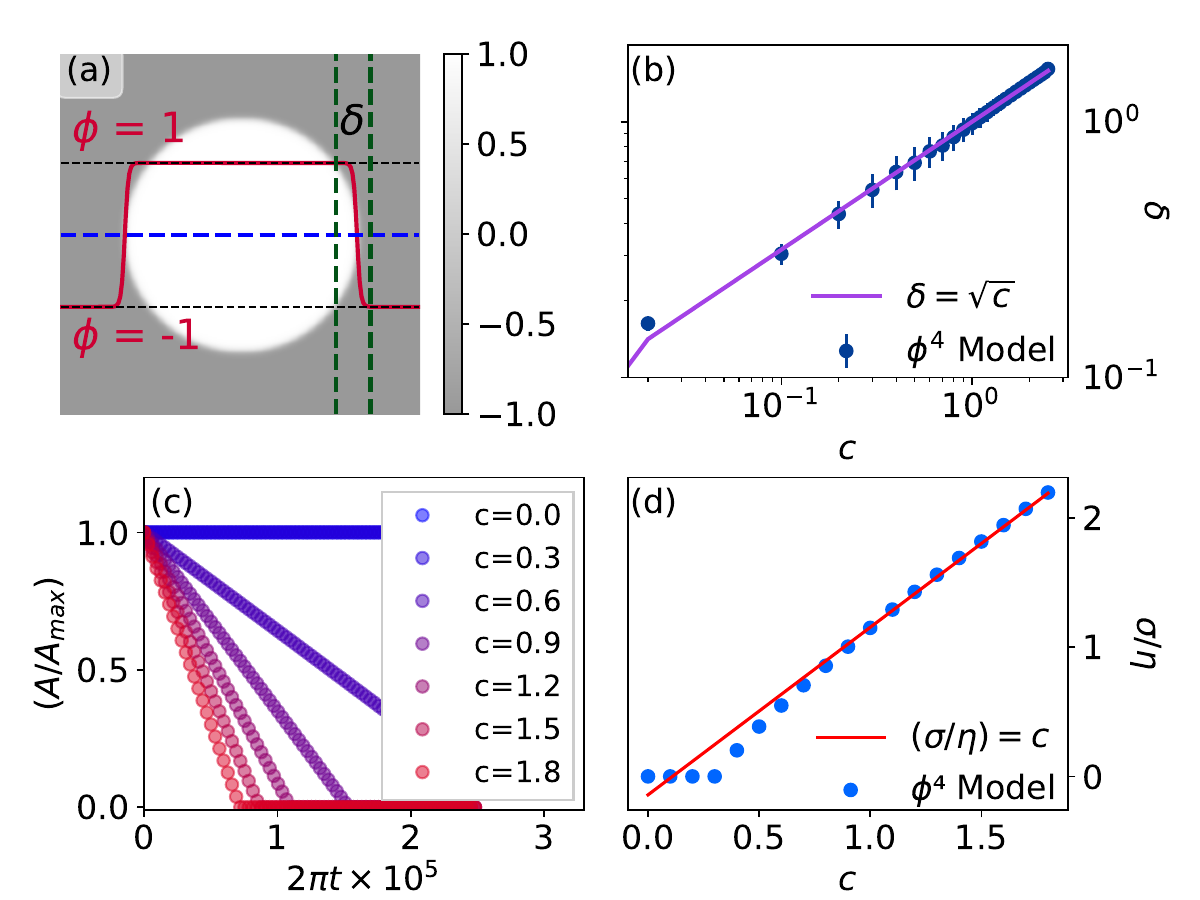}
  \caption{
  Numerical determination of the effective elastic tension constant $\sigma$ and the width $\delta$ of domain walls in the $\phi^4$ model. 
  (a) Example of a DW profile $\phi(x)$ at the center of a circular domain. The profile is fitted with a $\tanh(x/\delta)$ function to extract the wall width $\delta$. 
  (b) Dependence of the DW width $\delta$ on the stiffness parameter $c$. Symbols correspond to numerical simulations, while the solid line is the theoretical prediction $\delta \sim \sqrt{c}$. 
  (c) Evolution of the normalized area $A/A_{\max}$ as a function of $2 \pi t$ for different values of $c$, starting from an initial radius $R=90$. The slope of the decay increases with $c$. 
  (d) Dependence of the decay slopes obtained from panel (c) on $c$. Symbols correspond to simulations, while the solid line is a fit consistent with $\sigma/\eta \sim c/\gamma$.  
  These numerical results confirm the scaling laws $\delta \sim \sqrt{c}$ and $\sigma/\eta \sim c/\gamma$, \color{myblue} for large enough $c$ so to avoid mesh effects. \color{black}
  }
  \label{fig:VariosC}
\end{figure}

To obtain these relations numerically, we start from a circular domain that shrinks under curvature, in the abence of disorder. After a short transient, the DW profile $\phi(x)$ at the center of the domain is extracted and fitted with a $\tanh(x/\delta)$ function to determine $\delta$ (see Fig. \ref{fig:VariosC}(a)). Repeating for different values of $c$ we confirm $\delta \propto \sqrt{c}$ (see Fig. \ref{fig:VariosC}(b)).
Fitting the area evolution $A(t) = \pi R^2(t)$ versus $t$ for different values of $c$ (see Fig. \ref{fig:VariosC}(c)) yields $\sigma/\eta = k c / \gamma$, with $k \approx 1.14$ (see Fig. \ref{fig:VariosC}(d)).  

This constant $k$ enables quantitative comparison between analytical predictions for generic elastic loops and direct $\phi^4$ model simulations. The scaling $\delta \sim \sqrt{c}$ obtained numerically is consistent with the standard continuum field-theory prediction for domain-wall solutions of the $\phi^4$ model.

\section{DC-driven DWs in the $\phi^4$ model with quenched disorder}
\label{Phi4-Details}

\begin{figure}[h!]
  \centering
  \includegraphics[width=\linewidth]{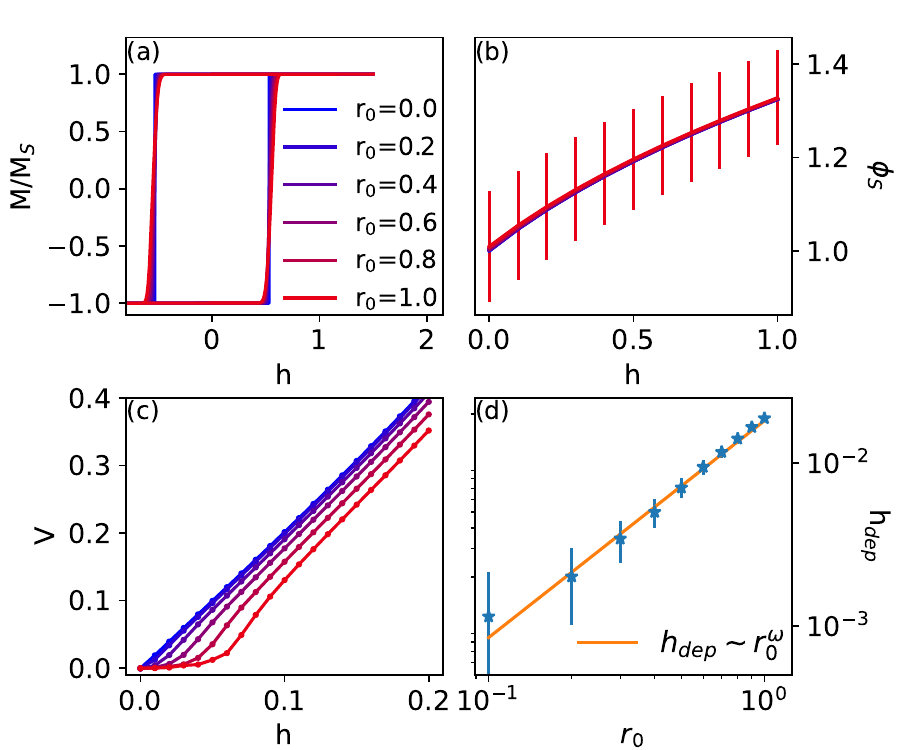}
  \caption{
  \color{black}
  Domain-wall dynamics in the $\phi^4$ model under a DC drive at $T=0$ with quenched disorder.
  (a) Hysteresis loops for different disorder strengths $r_0$. The coercive field and the saturation value $\phi_{\rm s}$ remain nearly independent of $r_0$ in the range studied.  
  (b) Dependence of $\phi_{\rm s}$ on the applied field $h$, in contrast with micromagnetic models where the saturation magnetization plateaus.  
  (c) Velocity--field characteristics for the same $r_0$ values, showing a continuous depinning transition at $h_{\rm d}(r_0)$.  
  (d) Scaling of the depinning field $h_{\rm d}$ with disorder strength, $h_{\rm d} \sim r_0^\omega$, $\omega\approx 1.33$, consistent with Ref.~\cite{kolton2023}.
  }
  \label{fig:Vel-Pared-Plana}
\end{figure}

We now characterize domain walls (DWs) in the $\phi^4$ model under a DC drive at zero temperature in the presence of quenched disorder. Figure~\ref{fig:Vel-Pared-Plana}(a) shows the hysteresis loop for different disorder strengths $r_0$. Both the coercive field and the saturation magnetization $\phi_{\rm s}$ remain practically independent of $r_0$ within the range explored. Nevertheless, as shown in Fig.~\ref{fig:Vel-Pared-Plana}(b), $\phi_{\rm s}$ still depends on the applied field $h$, unlike in true micromagnetic models where the saturation magnetization is strictly constant.  

Figure~\ref{fig:Vel-Pared-Plana}(c) presents the mean velocity-field characteristics as a function of the DC drive $h$ for the same disorder strengths $r_0$ used in panel (a). We observe a continuous depinning transition at $h_{\rm d} = h_{\rm d}(r_0)$, such that $V=0$ for $h<h_{\rm d}$ and $V>0$ for $h>h_{\rm d}$. The dependence of $h_{\rm d}$ on $r_0$ is shown in Fig.~\ref{fig:Vel-Pared-Plana}(d), where we find $h_{\rm d} \sim r_0^\omega$, with $\omega \approx 4/3$, in agreement with Ref.~\cite{kolton2023}.  

As discussed in Sec.~\ref{sec:AlternatingDriveSimulationsWithDisorder} and in Ref.~\cite{glatz2003}, the DC depinning transition of Figure~\ref{fig:Vel-Pared-Plana}(c) becomes rounded in the AC-driven case, when $V$ is the staggered mean velocity, $V=\langle v(t)\text{sgn}(h(t))\rangle$, with $v(t)$ the instantaneous center of mass DW velocity.

\section{Experimental Methods}
\label{sec:AppExperiments}

\begin{figure}[t]
  \centering
    \includegraphics[scale=0.55]{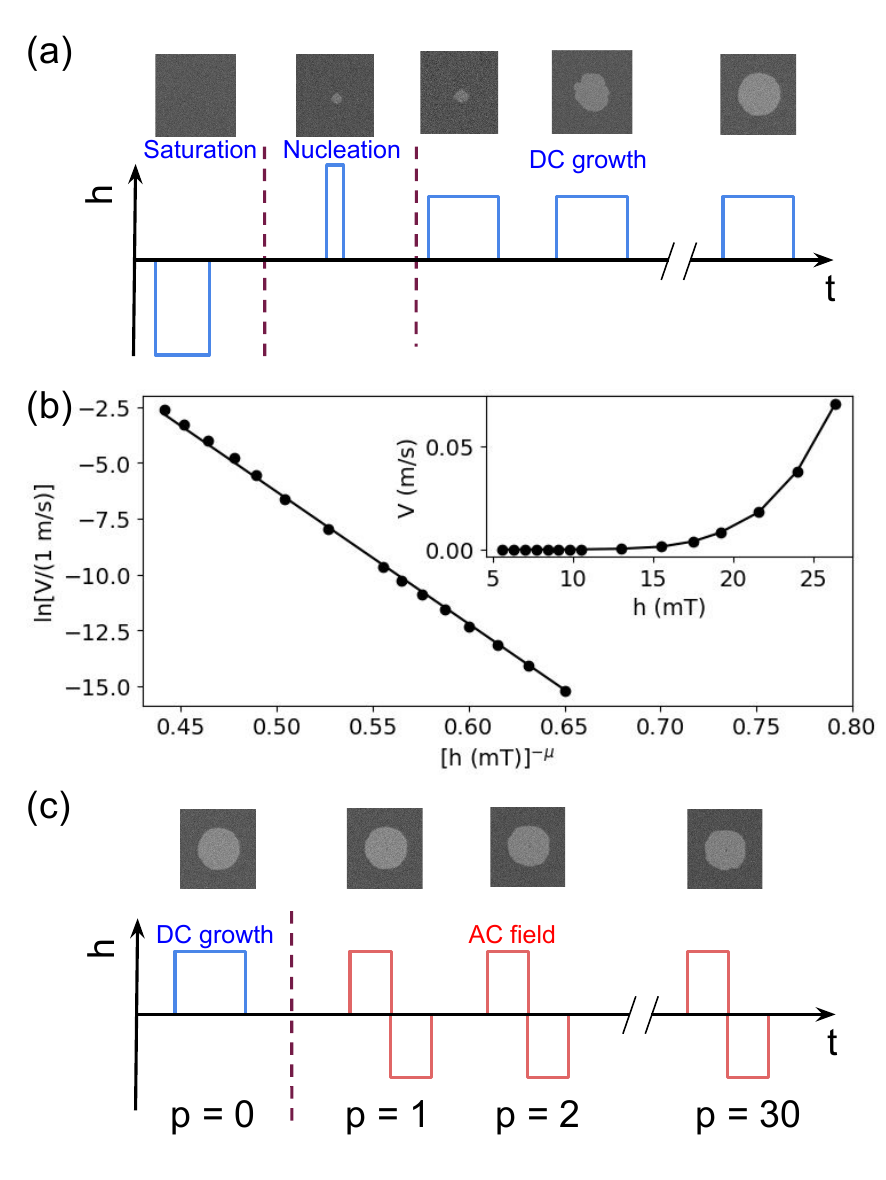}
    \caption{
    (a) DC protocol for domain growth: the procedure begins with full saturation of the sample using a -20~mT magnetic field pulse of 50~ms duration. A small domain is then nucleated with a pulse of similar amplitude and approximately 1~ms duration. Domain growth proceeds by applying 10–20~mT pulses in the same direction as the nucleation pulse. (b) Velocity results from DC growth experiments, showing $\ln[V/(1~\mathrm{m/s})]$ versus $h^{-\mu}$ with $\mu = 0.25$, for samples S1–S3, consistent with the creep law, (Inset: $V$ versus $h$). (c) AC protocol: following DC growth, zero-mean square magnetic field pulses are applied. After each pulse, a MOKE image is acquired, illustrated here with domain configurations from sample S5 at different cycle numbers $p$.
    }
    \label{fig:Protocolo}
\end{figure}

The AC experimental data of section Sec. \ref{sec:alternatingfieldexperiments}were obtained from an ultrathin ferromagnetic film with perpendicular magnetic anisotropy, all exhibiting bubble-type domains. The sample have the composition Ta(5)/Pt(3)/Co(0.8)/Ir(1)/Ta(2), where layer thicknesses are given in nanometers~\cite{domenichini2024}.
Magneto-Optical Kerr Effect (MOKE) microscopy was used to determine 
domain-wall velocities under constant fields (DC) and another for probing the response under alternating fields (AC).
The protocol for measuring DC and AC response is illustrated in Fig.~\ref{fig:Protocolo} ~\cite{domenichini2024}. In Fig. Fig.~\ref{fig:Protocolo}(b) we see that the sample accurately 
follows the creep law,  
\begin{equation}
v = \exp{(\alpha h^{-\mu} + \beta)},    
\end{equation}
where $\mu$ is the dynamic creep exponent ($\mu = 0.25$), and $\alpha$ and $\beta$ are fitting parameters 
for each sample ~\cite{domenichini2024}. 
We used this creep velocity 
to calculate ${\tilde \Lambda}$
for Fig.\ref{fig:acexperiments}.

\section{Samples Parameters}

The parameters of the ultrathin films [Co/Ni]$_4$, Pt/Co/Ir, and Pt/Co/Pt were obtained by combining data from Refs.~\cite{jeudy2018pinning, Domenichini_2019, domenichini2021, domenichini2024}.
In doing so, we used the relation
$$
T_d = \frac{\alpha H_{\rm d}^{4}}{T},
$$
with $T = 293\,\mathrm{K}$ in all cases, where $\alpha$ is obtained from fits of the creep law for each sample.
The parameters of the CoFeB thin films were subsequently obtained by fitting the decay of the semicircular domain radius shown in Sec.~\ref{sec:deflating}, while enforcing consistency with the values reported by Jeudy \emph{et al.}~\cite{jeudy2018}.

\begin{table}[t]
\centering
\caption{
Sample parameters for the data analyzed in Figs.~\ref{fig:lifetime} and \ref{fig:Rd-CoFeB}.
Shown are the depinning temperature $T_d$, depinning field $H_{\rm d}$, domain-wall surface tension $\sigma$, saturation magnetization 
$\phi_{\rm s}$, film thickness $t$, and characteristic  velocity $v_{\rm 0}$ (creep law prefactor).
The last three rows report derived quantities: 
$C = \sigma/(2 \phi_{\rm s})$, 
the depinning radius $R_{\rm d} = C/H_{\rm d}$, and the characteristic half-radius lifetime $\tau_{1/2}$ for an initial radius $R_0 = 10\,\mu\mathrm{m}$.}
\label{tab:samples}
\begin{tabular}{|l|c|c|c|c|}
\hline\hline
 & [Co/Ni]$_4$  & Pt/Co/Pt  & Pt/Co/Ir  & CoFeB \\
\hline
$T_d [K]$   & 14367 & 6533 & 4196 & 2230 \\
$H_{\rm d}$[mT]   & 12.5 & 56 & 74 & 7 \\
$\sigma$[mJ/m$^2$]  & 2.26 & 2.26 & 18 & 5.1 \\
\color{myblue}
$\phi_{\rm s}$[kA/m]  
\color{black}
& 540 & 910 & 1140 & 1000 \\
$t$[nm]    & 3.2 & 1 & 0.8 & 1 \\
$v_0$[m/s]   & 15.7 & 14 & 4.4 & 0.2 \\
\hline
$C$ [$10^{-6}$ mJ/(A\,m)]    & 2.1 & 1.2 & 7.9 & 2.5 \\
$R_{\rm d}$[$\mu$m]   & 0.17 & 0.02 & 0.11 & 0.36 \\
$\tau_{1/2}$ (s)         & $2.5 \times 10^{51}$  & $3.8 \times 10^{38}$  & $1.3 \times 10^{13}$  & $1.3 \times 10^{3}$  \\
\hline\hline
\end{tabular}
\end{table}
\color{black}


\bibliography{referencias}

\newpage
\tableofcontents
\setcounter{tocdepth}{2}

\end{document}